\documentclass[reprint,aps,nofootinbib,superscriptaddress,pra]{revtex4-2}

\usepackage{braket}
\usepackage{slashed}
\usepackage{graphicx}
\usepackage{xcolor}
\usepackage{enumitem}
\usepackage{siunitx}
\usepackage{mathtools}
\usepackage{dsfont}
\usepackage[hidelinks]{hyperref}
\usepackage{amsmath,amsthm,amsfonts,amscd,amssymb}

\newcommand{\tin}{\mathrm{in}}
\newcommand{\tout}{\mathrm{out}}

\newcommand\blfootnote[1]{%
  \begingroup
  \renewcommand\thefootnote{}\footnote{#1}%
  \addtocounter{footnote}{-1}%
  \endgroup
}

\usepackage{tikz} 
\usetikzlibrary{shapes,arrows,positioning,automata,backgrounds,calc,er,patterns}
\usetikzlibrary{decorations,decorations.markings,decorations.pathreplacing,decorations.pathmorphing}
\usepackage{relsize}
\usetikzlibrary{quantikz2}

\newcommand{\be}{\begin{equation}}
\newcommand{\ee}{\end{equation}} 
\newcommand{\mb}{\mathbf}

\begin{document}

\title{Impulse measurements enhanced with squeezed readout light}

\author{Tsai-Chen Lee}
\affiliation{Physics Division, Lawrence Berkeley National Laboratory, Berkeley, CA}
\affiliation{Department of Physics, University of California, Berkeley, CA}
\author{Jacob L. Beckey}
\affiliation{Department of Mathematics, University of Illinois Urbana-Champaign}
\affiliation{Illinois Quantum Information Science and Technology Center (IQUIST), University of Illinois Urbana-Champaign}
\author{Giacomo Marocco}
\affiliation{Physics Division, Lawrence Berkeley National Laboratory, Berkeley, CA}
\author{Daniel Carney}
\affiliation{Physics Division, Lawrence Berkeley National Laboratory, Berkeley, CA}

\begin{abstract}
    
We quantify how squeezed light can reduce quantum measurement noise to levels below the standard quantum limit in impulse measurements with mechanical detectors. The broadband nature of the signal implies that frequency-dependent squeezing performs better than frequency-independent squeezing. We calculate the optimal scaling of the impulse sensitivity with the squeezing strength, and quantify degradations due to photodetection losses. Even for lossless measurement, we find there exists a fundamental limit to the benefit of squeezing that depends only on the system's mechanical properties.

\end{abstract}

\maketitle

\blfootnote{tsaichen\_lee@berkeley.edu \\ ~jbeckey@illinois.edu \\ gmarocco@lbl.gov \\ carney@lbl.gov}

\section{Introduction}
It is increasingly common in modern tests of fundamental physics, like gravitational-wave detection \cite{harry2010advanced,somiya2012detector,acernese2014advanced} or dark matter searches \cite{brubaker2018first,dixit2021searching}, for quantum noise to be the dominant source of noise. This implies that one must engineer the quantum noise in order to mitigate its effects and increase the science reach of these quantum-limited experiments \cite{caves1981quantum,aasi2013enhanced, mcculler2020frequency-dependent,backes2021quantum,malnou2019squeezed, ghosh2022combining}. As such, the theory of quantum measurement has become increasingly relevant in fundamental physics in recent years~\cite{beckey2023quantum}.

Quantum noise reduction is particularly relevant for table-top experiments involving the optomechanical measurement of levitated sensors operating close to the quantum limit~\cite{ delic2020cooling, tebbenjohanns2021quantum, magrini2021Realtime, ranfagni2022two, kamba2022optical}. These systems have many exciting applications ranging from electric and gravitational field sensing~\cite{moore2014Search,Monteiro:2020qiz} to searches for new physics beyond the standard model~\cite{moore2021Searching,blakemore2021Search}. Ref.~\cite{gonzalez-ballestero2021levitodynamics} gives a thorough review of applications of levitated sensors, most of which operate by measuring the position of trapped particles, resulting in sensitive probes of forces or changes in  momentum~\cite{mason2019continuous, Monteiro:2020qiz,Monteiro:2020wcb,Carney:2022pku}. In this paper, we study the quantum limits that arise when using such set-ups to detect weak, approximately instantaneous impulses, and how to surpass them.

As we will see below, the relevant figure of merit when attempting to sense an instantaneous change in momentum is the \textit{momentum threshold}~\cite{clerk2004Quantumlimited,ghosh2020backaction}
\begin{align}
\label{eq:pth}
\Delta p = \left[ \int_{-\infty}^{\infty} \frac{d\nu}{2\pi \,S_{FF}(\nu)} \right]^{-1/2},
\end{align}
which is given in terms of the force power spectral density $S_{FF}(\nu)$ at angular frequency $\nu$. This threshold is a measure of the minimum resolvable impulse or momentum ``kick'', and is lower bounded $\Delta p \geq \sqrt{m \omega_m}$ for coherent input states.
Crucially, $\Delta p$ depends on the noise at all frequencies, which makes determining the benefit of using quantum-enhanced strategies, like squeezed light, non-trivial. If only frequency-independent squeezed light is used, the force power spectral density (PSD) will be lowered over some frequency bandwidth but will increase elsewhere. Thus, it is not, a priori, clear what benefit frequency-independent squeezed light has to offer in impulse sensing.

The most famous example of using squeezed light to enhance a fundamental physics experiment is LIGO, where techniques beyond frequency-independent squeezing have been considered. As was pointed out theoretically decades ago~\cite{caves1981quantum,unruh1983quantum,kimble2001conversion} and demonstrated experimentally recently in cavity-based gravitational wave detectors~\cite{mcculler2020frequency-dependent,ganapathy2023broadband}, to best reduce noise in a broadband way with squeezed light, the quadrature of light whose variance should be squeezed may vary over the detection bandwidth. 

For an approximately free mass, such \textit{frequency-dependent} squeezed light always results in a reduced force PSD. Far from a mechanical resonance, the force PSD can be exponentially suppressed by a factor $S_{FF} \to e^{-2r} S_{FF}$ compared to using a coherent state. Here, $r$ quantifies the squeezing strength, typically defined by a reduction in the laser shot noise, i.e., a reduction in the show noise power $S_{YY} \to e^{-2r} S_{YY}$. In terms of the impact on the momentum threshold, by inspection of Eq.~\eqref{eq:pth}, one might expect that with frequency-dependent squeezed light the threshold could scale as
\begin{align}
\label{dp-expected}
    \Delta p \rightarrow e^{-r}\, \Delta p.
\end{align}
However, for optomechanical systems which feature a mechanical resonance within the detection bandwidth (as in several modern experiments involving levitated dielectrics~\cite{delic2020cooling,wang2024Mechanical}), the noise on-resonance cannot be reduced below that achievable with coherent states, and so deriving the scaling of $\Delta p$ with $r$ is in principle more complicated. 
 
In this paper, we quantify the effects of squeezed light on the momentum threshold for mechanical oscillators that exhibit resonances, and show that improvements to the threshold $\Delta p$ can approach the ideal $e^{-r}$ scaling of Eq. \eqref{dp-expected}. We also quantify two limitations to this idealized scaling. Firstly, we find a fundamental limit set by the the finite damping rate $\gamma = \omega_m/Q$ of the oscillator, such that even with an arbitrary amount of squeezing and perfect photodection, the lowest possible threshold is given by $\Delta p = \Delta p_{\rm SQL}/\sqrt{Q}$. The second comes from optical loss: with a finite photodetection efficiency $\eta < 1$, losses degrade the benefits of squeezed light, although sub-SQL thresholds are still achievable. At small efficiencies, we show that the scaling changes to $\Delta p \sim e^{-r/2}$.

Working towards an understanding of modern experiments which involve dielectric nano-particles trapped in three dimensions~\cite{gonzalez-ballestero2021levitodynamics, delic2020cooling, tebbenjohanns2021quantum, magrini2021Realtime, ranfagni2022two, kamba2022optical,moore2021Searching}, we consider two different experimental set-ups that could be used to detect small impulses. First, we consider the canonical example of a cavity with a moveable mirror, i.e., a Fabry-P\'erot cavity. Then, moving towards a model that captures more of the physics involved in trapped-particle detectors, we consider a harmonically suspended dielectric slab in free space. This dielectric slab model contains the same underlying interaction Hamiltonian as a dielectric nanosphere, but with the simplifying constraint that the scattering of light is purely one-dimensional, allowing a more transparent theoretical treatment. We show that the quantum noise properties of the dielectric slab may be obtained as the limit of cavity optomechanics where the cavity damping rate is much larger than the other frequency scales of the system, indicating that the optimal measurement strategies are the same in both cases.

Because of this equivalence, our simulations are based on the dielectric slab because it is closer to the actual experiments involving suspended spherical dielectrics. In an effort to inform current experiments, we focused on a parameter regime similar to those in Ref.~\cite{magrini2021Realtime}. The parameters used in our simulations are summarized in Table~\ref{tab:sphere-params}.

\setlength{\tabcolsep}{12pt}
\begin{table}
\centering
\renewcommand{\arraystretch}{1.3} 
\begin{tabular}{ c c c }
 Parameter & Symbol & Value \\ 
 \hline
 mass & $m$ & $10^{-18}~\text{kg}$ \\
 mechanical frequency & $\omega_m/2\pi$ & $10^5~{\rm Hz}$ \\
 damping rate & $\gamma/2\pi$ & $10~{\rm Hz}$ \\
 polarizability & $\chi_\mathrm{e}$ & 3.5 \\
  laser wavelength & $\lambda$ & $1500$ nm \\
 effective temperature & $T$ & $200 \times 10^{-6}~{\rm K}$  \\
\end{tabular}
\caption{\textbf{Dielectric slab parameters:} A summary of the parameters used in our model. These parameters are representative of cutting-edge experiments involving levitated dielectric spheres (see for example Ref.~\cite{magrini2021Realtime}).}
\label{tab:sphere-params}
\end{table}

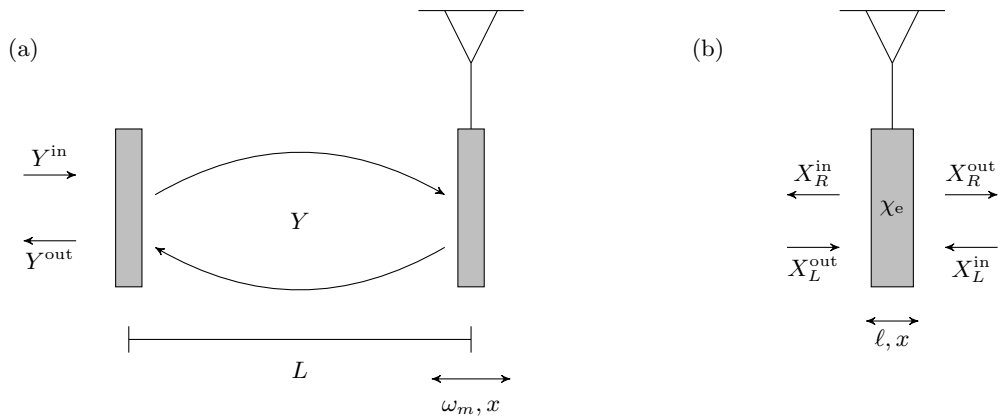
\begin{figure*}[ht] 
    \centering
    \begin{tikzpicture}[scale=0.35, >=stealth', pos=.8, photon/.style={decorate, decoration={snake, post length=1mm}}]

        \draw[->] (-19,1) to[out=30, in=150] (-8,1);
        \draw[->,yscale=-1] (-8,1) to[out=150, in=30](-19,1);

        \node at (-13.5,0) {$Y$};
        \draw[->] (-24, 1.75) -- (-22, 1.75) node[midway, above] {$Y^{\mathrm{in}}$};
        \draw[->] (-22, -0.75) -- (-24, -0.75) node[midway, below] {$Y^{\mathrm{out}}$};
        
        \draw [black, fill=lightgray] (-19.5, -2.5) rectangle (-20.5, 3.5);
        \draw [black, fill=lightgray] (-6.5, -2.5) rectangle (-7.5, 3.5);
        
        \draw (-7, 3.5) -- (-7, 6);
        \draw (-5, 8) -- (-9, 8);
        \draw (-7, 6) -- (-6, 8);
        \draw (-7, 6) -- (-8, 8);
        
        \draw[<->] (-8.5, -6) -- (-5.5, -6) node[midway, below=5pt] {$\omega_m, x$};
        \draw (-20, -4.5) -- (-7, -4.5) node[midway, below=5pt] {$L$};
        \draw (-20, -5) -- (-20, -4) ;
        \draw (-7, -5) -- (-7, -4) ;

        \node at (-24, 6.5) {(a)};
        
        \draw [black, fill=lightgray] (9.8, -2.5) rectangle (8.2, 3.5) node[midway] {$\chi_\mathrm{e}$};
        \draw (9, 3.5) -- (9, 6);
        \draw (7, 8) -- (11, 8);
        \draw (9, 6) -- (8, 8);
        \draw (9, 6) -- (10, 8);
        
        \draw[->] (11, 1) -- (13, 1) node[midway, above] {$X_{R}^{\mathrm{out}}$};
        \draw[->] (7, 1) -- (5, 1) node[midway, above] {$X_{R}^{\mathrm{in}}$};
        \draw[->] (5, -1) -- (7, -1) node[midway, below] {$X_{L}^{\mathrm{out}}$};
        \draw[<-] (11, -1) -- (13, -1) node[midway, below] {$X_{L}^{\mathrm{in}}$};
        
        \draw[<->] (10, -3.8) -- (8, -3.8) node[midway, below] {$\ell, x$};

        \node at (2, 6.5) {(b)};
        
    \end{tikzpicture}
    
    \caption{(a) A canonical example of an optomechanical sensor is the Fabry-P\'erot cavity -- comprised of two parallel mirrors, one partially transmissive and one highly reflective and treated as a mechanical oscillator with mass $m$ and resonance frequency $\omega_m$. (b) Schematic diagram of a dielectric slab of mass $m$, width $\ell$, and electric susceptibility $\chi_\mathrm{e}$ suspended as a harmonic pendulum with frequency $\omega_m$, interacting with left- and right-moving electromagnetic waves. The coordinate $x$ denotes the displacement of the slab from its classical equilibrium position. }
    \label{fig:FP-and-slab}
\end{figure*}

\section{Optomechanics background}

The field of levitated optomechanics has undergone substantial advancement in the last 50 years, with particles ranging from $50$~nm to $10~\mu$m being successfully trapped using various techniques~\cite{millen2020Optomechanics}. Because these systems can be substantially isolated from environmental effects when operated in high-vacuum, they have great potential as quantum sensors~\cite{millen2020Optomechanics,gonzalez-ballestero2021levitodynamics,moore2021Searching}. While modern experiments involving levitated particles, of course, contain a plethora of subtle experimental and theoretical details, the basic working principle can be understood by studying a simple Fabry-Perot cavity or a harmonically suspended dielectric slab (as shown in shown in Fig.~\ref{fig:FP-and-slab}).  

We assume that these dielectrics are suspended via a harmonic potential and then interrogated by an electromagnetic field to obtain information about the time-dependent position, $x(t)$. From this data, one can infer information about impulses using linear response theory. Let us briefly review the formalism of how this works (see, for example Sec. IIC of Ref~\cite{beckey2023quantum} for more details).

\subsection{Input-output formalism}

The input-output formalism, as the name suggests, relates output fields to input ones, and models noise sources as oscillator baths that couple to the system of interest~\cite{bowen2016quantum}. As long as this coupling is linear, the resultant input and output operators, ($O_{i}^{\rm in}$ and $O_{i}^{\rm out}$, respectively) will be related linearly in the frequency domain as
\begin{align}
    O_i^{\rm out} (\nu) &= \sum_j \chi_{ij}(\nu) O_j^{\rm out} (\nu),
\end{align}
where the $\chi_{ij}(\nu)$ are susceptibilities. In this work, we will focus on the case where our observable is the phase quadrature of light, denoted $Y^{\rm out}$. Thus, the input-output relation will take the form (suppressing the frequency dependence to ease notation)
\begin{align}
    Y^{\rm out} &= \chi_{YY} Y^{\rm in} + \chi_{YX} X^{\rm in} + \chi_{YF} F^{\rm in},
    \label{eqn:generalOut}
\end{align}
where the specific forms of the susceptibilities are given in \eqref{eq:transfer-OM} and \eqref{eqn:slabChis}.

Throughout this paper, we consider two physical systems, depicted in Fig.~\ref{fig:FP-and-slab}. The first one is a Fabry-P\'erot cavity, predominantly in the bad-cavity limit where the cavity loss rate $\kappa$ is much larger than the mechanical frequency $\omega_m$. The second one is a dielectric slab which is a planar-symmetric version of a 3D levitated nanosphere, whose high degree of symmetry simplifies the treatment while retaining the core physics of the measurement. As we show in  App.~\ref{app:slab-details}, the output phase quadrature for a bad cavity and the left-moving output phase quadrature for a dielectric slab take the same functional form (suppressing frequency dependence to ease notation),
\begin{align}
    Y_\mathrm{cav}^\tout &\approx - Y^\tin - \frac{8 g_c^2}{\kappa} \chi_m X^\tin + \sqrt{8 g_c^2}{\kappa} \chi_m F^\tin, \\
    Y_{L,\mathrm{slab}}^\tout &= Y_L^\tin  - 2 g^2 \chi_m X_R^\tin + \sqrt{2} g\chi_m F^\tin,
    \label{eqn:outputPhaseEq}
\end{align}
where $\chi_m(\nu) = [m(\omega_m^2 - \nu^2 + i \gamma \nu)]^{-1}$ is the mechanical susceptibility that depends on the mechanical mass $m$ and the damping rate $\gamma$, and $g_c$ and $g$ are the drive-enhanced, optomechanical couplings for the cavity and a diectric slab, respectively. Since these two output quadratures are of the same functional form, conclusions we reach regarding one system may be directly applied to the other with a replacement of constants. The correspondence is spelled out in \ref{app:map}. We will make use of this correspondence throughout the remainder of the paper.

For many sensing applications, one wants to infer information about the force from this output quadrature data. To do so, we take an estimator of the form
\begin{align}
    F_E(\nu) = \frac{Y^{\rm out}(\nu)}{\chi_{YF}(\nu)},
\end{align}
where $Y^\mathrm{out} (\nu)$ is given by \eqref{eqn:outputPhaseEq} for the systems of interest. It then remains to quantify how sensitive a force one can estimate. This is done by computing the force \textit{power spectral density} (PSD), $S_{FF}(\nu)$, which quantifies the force noise per unit frequency in our estimator and is obtained assuming stationary noise via
\begin{align}
    2\pi S_{FF} (\nu) \delta(\nu - \nu') &= \langle F_E(\nu) F_E(\nu') \rangle.
    \label{eqn:PSDdef}
\end{align}
For linear systems, for which the output phase is given as in \eqref{eqn:generalOut}, we have
\begin{align}
\begin{split}
\label{eq:SYY-OM}
S_{FF} = & \frac{|\chi_{YY}|^2}{|\chi_{YF}|^2} S^{\rm in}_{YY} + \frac{|\chi_{YX}|^2}{|\chi_{YF}|^2} S^{\rm in}_{XX}  \\
& + \frac{2 \mathbf{Re}\{\chi_{YX} \chi^*_{YY}\}}{|\chi_{YF}|^2} S^{\rm in}_{XY} + S_{FF}^\tin,
\end{split}
\end{align}
which plays a central role throughout our work. We show some examples of force PSDs in Fig.~\ref{fig:coherentPSD}(a). Among the terms that contribute to these PSDs, there is $S_{YY}^{\rm in}$ ($S_{XX}^{\rm in}$), due to the quantum fluctuations in the input phase (amplitude) quadrature, while the $S_{FF}^{\rm in}$ term represents environmental noise sources like collisions with ambient gas particles. We assume that the system is dominated by quantum noise, and so we shall set \begin{align}
S_{FF}^{\rm in}= m\gamma \nu
\label{eqn:main_QN}
\end{align} throughout, corresponding to the minimal noise arising from quantization of the test mass (see App. \ref{app:FD_theorem} for a derivation of this). The remaining so-called cross-term depends on $S_{XY}$, which vanishes for coherent states. However, when the injected light is squeezed or non-Gaussian, the cross-term can be made negative, thus lowering the overall noise over some squeezing bandwidth. We give explicit expressions for these quantities in terms of squeezing parameters in \eqref{eqn:SXX_squeezed}.

While the force PSD at a particular frequency is not the figure of merit one aims to optimize when detecting an instantaneous impulsive signal, it does enter calculation of the momentum threshold \eqref{eq:pth}. When all systematics and classical environmental noise sources have been mitigated, one is fundamentally limited by the quantum noise intrinsic in the light used to interrogate the mechanical system. Thus, to enhance the sensitivity of these detectors further, one must engineer the quantum noise itself. Before we discuss these quantum limits and how to surpass them, let us review the essentials of sensing impulsive signals.

\subsection{Impulse sensing}
\label{sec:coherentState}

The sensing task we are considering is the detection of particle-like events that occur infrequently within some long observation time $T_{\mathrm{int}}$. Thus, the signal is zero for almost all of $T_{\mathrm{int}}$, with only noise acting. The task becomes resolving a small number of distinct clicks modeled as delta function spikes of the form
\begin{align}\label{eq:Fsig}
    F_{\mathrm{sig}} = \Delta p_\mathrm{sig} \delta(t).
\end{align}
The best strategy to extract this signal turns out to be weighting the signal in the frequency domain by its noise. This is called a matching filter search, or ``template matching''~\cite{brunelli1997template, owen1999matched}, which we describe briefly before deriving the momentum threshold. For more details see Appendix B of Ref.~\cite{ghosh2020backaction}.

Suppose the data obtained from some measurement is represented by $F(t)$. To filter this data in order to detect an instantaneous impulse, $F_{\mathrm{sig}}$, occurring at time $t_e$, one introduces a so-called filter function $f(t-t_e)$ such that our estimator can be expressed
\begin{align}
    O(t_e) &= \int f(t-t_e) F (t) dt.
\end{align}
Given an estimator of this form, the goal is to then optimize over filter functions to maximize the signal-to-noise ratio. Doing so for a signal of the form in~\eqref{eq:Fsig}, one finds the optimal filter\cite{ghosh2020backaction}
\begin{align}
    f_{\mathrm{opt}} (\nu) &= \frac{\Delta p_\mathrm{sig}}{S_{FF}(\nu)},
\end{align}
leading to an upper bound on the signal-to-noise ratio (SNR) of
\begin{align}
{\rm SNR}_{\rm opt} =  \sqrt{\int_{-\infty}^{\infty} d\nu \frac{(\Delta p_\mathrm{sig})^2}{2 \pi \, S_{FF}(\nu)}}.
\label{eq:SNR-opt}
\end{align}
Demanding that this SNR be greater than or equal to unity yields the minimal resolvable $\Delta p_\mathrm{sig}$ given in Eq.~\eqref{eq:pth}
\begin{align}
 \Delta p_\mathrm{sig} \geq \Delta p = \left[ \int_{-\infty}^{\infty} \frac{d\nu}{ 2\pi S_{FF}(\nu)} \right]^{-1/2},
\label{eqn:threshold}
\end{align}
which gives a measure of the minimum detectable impulse.
We see that the momentum threshold in instantaneous impulse sensing depends upon the frequency integral of the inverse force PSD.

To warm up for the more complicated case of squeezed light, let us review how the Standard Quantum Limit for impulse sensing is derived. In order to evaluate the momentum threshold, we need to know the functional forms of the force PSD. These are derived from the solutions for the output phase quadrature \eqref{eqn:outputPhaseEq} and the definition of the PSD in \eqref{eqn:PSDdef}. Assuming coherent input light with no cross-correlations $S_{XY}=0$, the force PSD for a cavity in the bad cavity ($\kappa \gg \omega_m$) limit is
\begin{equation}
S^0_{F F}(\nu)=\frac{m^2 \kappa\left[\left(\nu^2-\omega_m^2\right)^2+\gamma^2 \nu^2\right]}{16 g_c^2}+\frac{4 g_c^2}{\kappa},
\end{equation}
where the first term is the ``shot-noise" and the second term is the ``back-action".
The PSD for a dielectric slab is identical up to its numerical coefficients:
\begin{equation}
S^0_{F F}(\nu) = \frac{m^2}{4 g^2} \left[\left(\nu^2-\omega_m^2\right)^2+\gamma^2 \nu^2\right] + g^2.
\label{eqn:coherentPSD}
\end{equation}
This similarity is a consequence of the similar bilinear coupling between the light modes and the oscillator's position in the underlying interaction Hamiltonians.

\begin{figure*}[ht!]
    \centering
    \includegraphics[width=\linewidth]{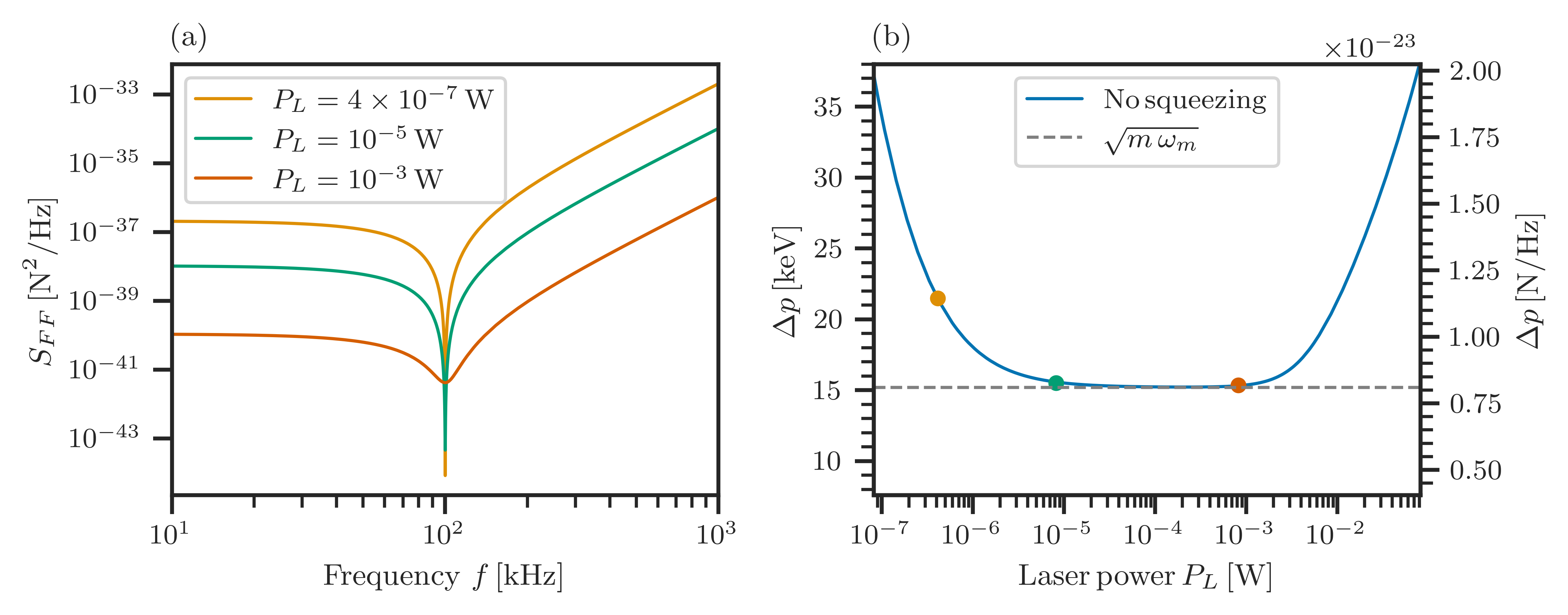}
    \caption{(a) The contribution of quantum measurement noise to the force PSD as given in \eqref{eqn:coherentPSD}, assuming vacuum-limited precision, for three different optomechanical couplings -- the laser power for each of three curves corresponds to the three colored dots in (b), where we see that the corresponding impulse sensitivities differ by at most a factor of $2^{1/2}$. (b) The momentum threshold as a function of loser power -- the power is related to the coupling $g$ through \eqref{eqn:power}. The plateau indicates there exists a range of powers with equal impulse sensitivities, where on- and off-resonance sensitivity can be exchanged for equivalent SNRs.  }
    \label{fig:coherentPSD}
\end{figure*}
Making use of these PSDs, we can evaluate the momentum sensitivity \eqref{eqn:threshold} to find the optimal impulse threshold with coherent input states~\cite{clerk2004Quantumlimited}. The integral over frequency may be written in terms of the dimensionless variable $x = \nu/\omega_m$ as
\begin{align}
    \Delta p = \Big[I_0 \int_0^\infty \frac{1}{(x^2-1)^2 + \frac{1}{Q^2}x^2 + a^2 + bx/Q}\Big]^{-1/2},
    \label{eq:threshold-sql}
\end{align}
where the mechanical quality factor $Q\equiv \frac{\omega_m}{\gamma} $. For a dielectric slab $I_0 = \tfrac{4g^2}{\pi m^2\omega_m^3}$, $a = \tfrac{2g^2}{m\omega_m^2}$, and $b=4g^2/m\omega_m^2$ (the bad cavity result is obtained with $g^2\to 4g_c^2/\kappa$).\footnote{This is a concrete example of a more general analysis in \cite{clerk2004Quantumlimited}, where the same integral is found and the condition $\Lambda \ll 1$ and $\Gamma \ll 1$ is required to obtain the SQL threshold $\sqrt{ m \omega_m}$. While Eq. \eqref{eq:threshold-sql} can be evaluated without the approximation, the lower bound of the threshold appears in this limit. Whether the system lives in this regime will depend on the mechanical quality factor Q and the coupling.} In a high-$Q$ mechanical system where the back-action is sufficiently small, i.e. $a \ll 1$, the integral is dominated by the resonance at $x=1$ and a large Q expansion shows
\begin{align}
    \Delta p = \left(\frac{g^2 + g_*(\omega_m)^2}{g^2}\right)^{1/2}
 \sqrt{m \omega_m},
 \label{eqn:pth(g)}
 \end{align}
up to small $1/Q$ corrections. Here, $g_*(\omega_m)$ is the coupling that mimimizes the PSD on-resonance \begin{align}
    \frac{\partial S^0_{FF}(\nu)}{\partial g} \Big|_{g = g_*} = 0,
    \label{eqn:gStar}
\end{align} i.e. $g_*(\omega_m) = \sqrt{m\gamma\omega_m/2}$ for the dielectric slab. Therefore, we prove that the momentum threshold is bounded by the ``Standard Quantum Limit" (SQL) 
\begin{align}
     \Delta p _\mathrm{SQL} =  \sqrt{m \omega_m},
    \label{eqn:thSQL}
\end{align}
which is saturated for large couplings satisfying $g_*(\omega_m) \ll g \ll \sqrt{Q} \times g_*(\omega_m)$. A comparison of \eqref{eqn:pth(g)} and \eqref{eqn:thSQL} demonstrates that setting $g = g_*(\omega_m)$ increases the momentum threshold only by a factor $2^{1/2} \approx 1.41$, independent of the mass and mechanical frequency of the trap. This modest decrease in sensitivity allows for much lower laser powers to be used.

The optomechanical coupling $g$ that appears in \eqref{eqn:pth(g)} is an experimentally tunable parameter, which is controlled by the laser power. For a dielectric slab, the relation between the (drive-enhanced) optomechanical coupling $g$ and the laser power $P_L$ is derived in \eqref{eqn:coupling} to be 
\begin{align}
   g \approx \frac{ \sqrt{\omega_0 P} \chi_\mathrm{e} \ell \omega_0}{2\pi} ,
    \label{eqn:power}
\end{align}
where $\omega_0$ is the laser's angular frequency and $\chi_\mathrm{e}$ is the electric suseptibility of the dielectric, and we have assumed $\ell \omega_0 \ll 1$. Using the parameters of Table \ref{tab:sphere-params}, we find the power necessary to achieve a coupling $g_*(\omega_m)$ is
$    P_L\big(g_*(\omega_m)\big)~\approx~\SI{4.4e-7}{W}.
$ We see in Fig.~\ref{fig:coherentPSD}(b) that once the laser power exceeds this value the dependence on the power is rather flat, and so increasing the benefit of further increasing the power is marginal.

The above results describe the optimal impulse sensitivity that can be achieved for coherent states -- in the absence of squeezing -- for the ideal case of a lossless system. In the next section, we investigate how the impulse sensitivity may be improved by the use of squeezed input light. 

\section{Impact of squeezing}

The force PSDs in Eq.~\eqref{eqn:PSDdef} for both a dielectric slab and a cavity can be altered by the use of input squeezed light. In the absence of squeezing the $S_{XY}^{\rm in}$ goes to zero and the force PSD becomes a simple trade-off between shot noise and radiation pressure, as we saw in the previous section. By using squeezed light with cross-correlations so that $S_{XY} \neq 0$, one has access to noise reduction strategies that are not available using coherent states. We now consider how frequency-independent input light can improve the momentum threshold, before turning to frequency-dependent squeezing.

\begin{figure*}[ht!]
    \centering
    \includegraphics[width=\textwidth]{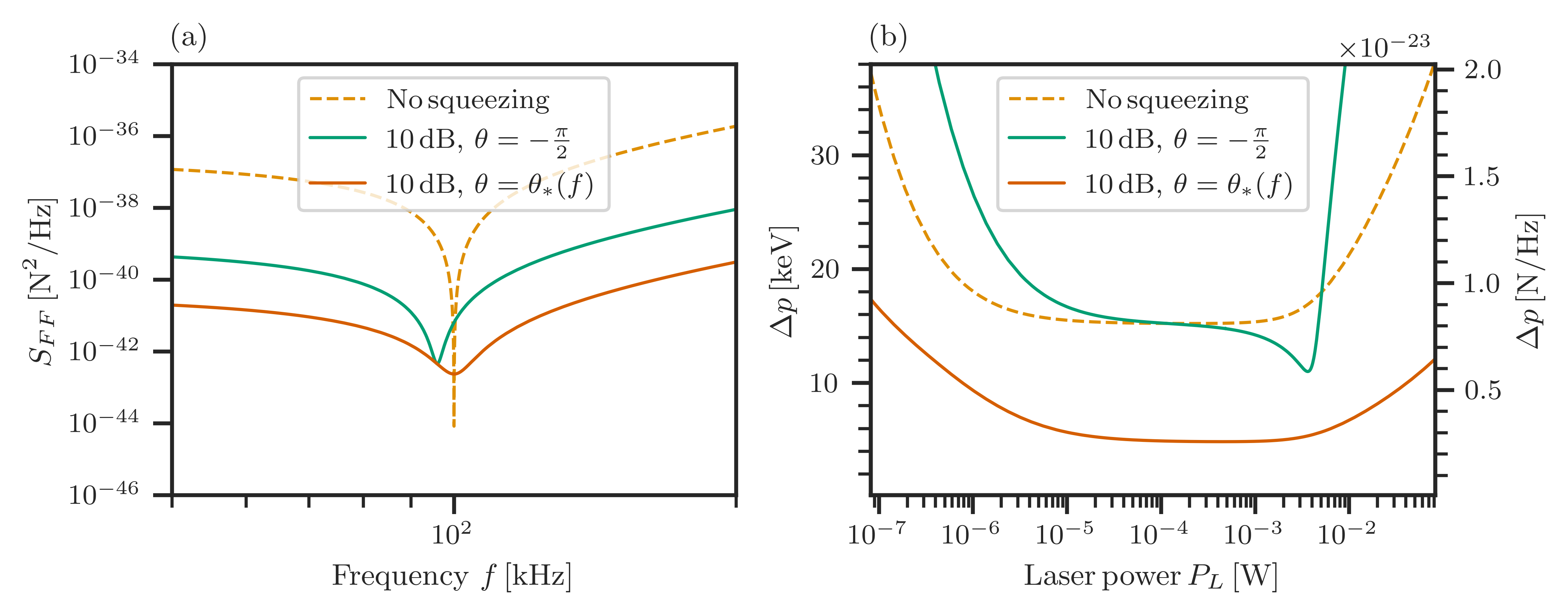}
    \caption{(a) The force PSD for a dielectric slab without squeezing, with 10 dB of frequency-independent squeezing with $\theta = -\frac{\pi}{2}$, and with 10 dB of frequency-dependent squeezing. In the latter two cases, we use a laser power of around 3 mW, which minimizes the frequency-independent threshold, while for the `No squeezing' case, we use $g=g_*(\omega_m)$. (b) The momentum threshold as a function of the laser power for the same three cases as (a). We see that the frequency-independent threshold performs worse the frequency-dependent threshold for equal squeezing strengths. }
    \label{fig:SFF-to-threshold}
\end{figure*}
 
\subsection{Frequency-independent squeezing}

Since the impulses we are interested in sensing arise from quasi-instantaneous momentum kicks, the size of the signal is approximately flat in frequency space. As such, the momentum threshold given in Eq.~\eqref{eq:pth} involves an integral over all frequencies, and so it is not immediately clear that the threshold SQL can be surpassed using frequency-independent (FI) squeezing. This is due to the fact that FI squeezing decreases the noise at some frequencies, while increasing it at others. The relationship between optimal laser power, squeezing strength, and squeezing angle is quite complicated, making analytical results regarding the use of FI squeezed light intractable. 
We can, however, numerically integrate Eq.~\eqref{eq:pth} for a given squeezing angle, power and  squeezing strength. 

Recall that the threshold depends on the integral of the \textit{inverse} of $S_{FF}(\nu)$. One can try to optimize the PSD at some particular frequency, and see how this affects the threshold. For a frequency-independent squeezing angle of $\theta = -\pi/2$, we see in Fig.~\ref{fig:SFF-to-threshold}(a) that this decreases the noise everywhere except around the resonance, where the noise is increased. By contrast, we also show the PSD achievable with frequency-dependent squeezing (to be examined more closely in the next section), which we see outperforms the frequency-independent strategy everywhere, as expected. Since Fig.~\ref{fig:SFF-to-threshold}(b) shows that for an equal squeezing strength, the optimal sensitivity is much better for frequency-dependent squeezing than frequency-independent squeezing, we shall focus for the remainder of the paper on the more effective strategy.

\subsection{Frequency-dependent squeezing}
\label{sec:FD_squeezing}

As discussed in the introduction, if one can generate squeezed light with the appropriate frequency dependence, the force PSD can be lowered at all frequencies ($S_{FF} \rightarrow e^{-2r}S_{FF}$) away from the mechanical frequency, suggesting that frequency-dependent (FD) squeezing may be better than FI squeezing. The idea of generating squeezed light with these properties was first proposed in Refs.~\cite{caves1981quantum,unruh1983quantum} and later worked out in detail in Ref.~\cite{kimble2001conversion}. They showed that if FI squeezed light was passed through a sequence of Fabry-P\'erot cavities detuned from the main laser frequency, one could engineer FD squeezed light. This proposal has now been implemented across a broad range of frequencies~\cite{chelkowski2005Experimental,mcculler2020frequency-dependent} and was recently implemented in LIGO itself~\cite{ganapathy2023broadband}. 

More recent proposals have also shown how to generate FD squeezed light using detuned optical parametric amplifiers~\cite{junker2022FrequencyDependent} or EPR-entangled beams~\cite{ma2017proposal,khalili2018Overcoming,zeuthen2019Gravitational}, providing multiple methods for potentially realizing these squeezing protocols. Regardless of the physical implementation, FD squeezing can be used to lower the force PSD at all frequencies, thus lowering the threshold.

\begin{figure}[t]
\includegraphics[width = \linewidth]{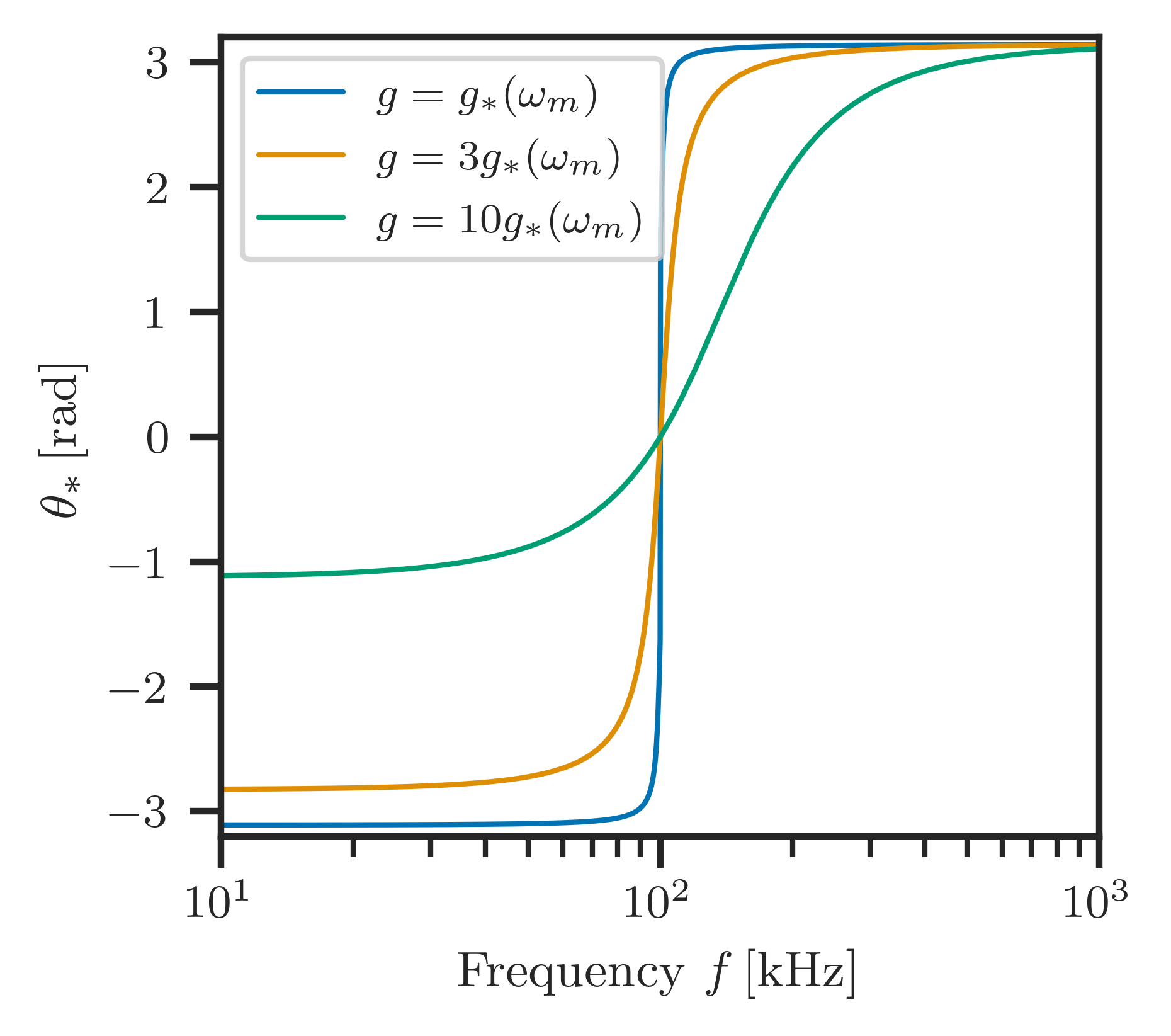}
    \caption{The optimal squeezing angle $\theta_*$, given in Eq.~\eqref{eqn:optimal_angle} that minimises the force PSD $S_{FF}(\nu)$ as a function of frequency $f = \nu/2\pi$. We show the results for increasing values of the optomechanical coupling $g$ in terms of $g_*(\omega_m) = \sqrt{m\gamma\omega_m/2}$, defined in \eqref{eqn:gStar}, that minimizes the coherent state PSD on-resonance.}
    \label{fig:optimal_angle}
\end{figure}

Concretely, we wish to use FD squeezing to minimize the PSD of Eq.~\eqref{eq:SYY-OM} by appropriately choosing the squeezing angle $\theta(\nu)$~\cite{kimble2001conversion}, which we emphasize may depend on frequency. By demanding $\partial_\theta S_{FF}(\nu) = 0$, we find the optimal angle is 
\begin{align}
    \tan \theta_*(\nu) = \frac{2 \mathrm{Re} \big[ \chi_{YX} \, \chi_{YY}^\ast \big]}{|\chi_{YX}|^2 - |\chi_{YY}|^2}
    \label{eqn:optimal_angle},
\end{align}
which we plot in Fig.~\ref{fig:optimal_angle}. Unless otherwise stated, the numerical plots and results in the main text are calculated with the dielectric slab.

Inserting Eq.~\eqref{eqn:optimal_angle} into the force PSD Eq.~\eqref{eq:SYY-OM}, we find
\begin{align}
    S_{FF}(\nu; r, \theta_*) = &\frac{1}{2|\chi_{YF}|^{2}} \Big[ \left( |\chi_{YX}|^2 + |\chi_{YY}|^2 \right) \cosh 2r \nonumber\\
    &- |\chi_{YX}^2 + \chi_{YY}^2 | \sinh 2r\Big] + m \gamma \nu,
    \label{eq:SFFopt}
\end{align}
which gives us the minimal noise at an angular frequency $\nu$ as a function of the squeezing strength $r$.

\begin{figure*}[ht!]
    \centering
    \includegraphics[width = \linewidth]{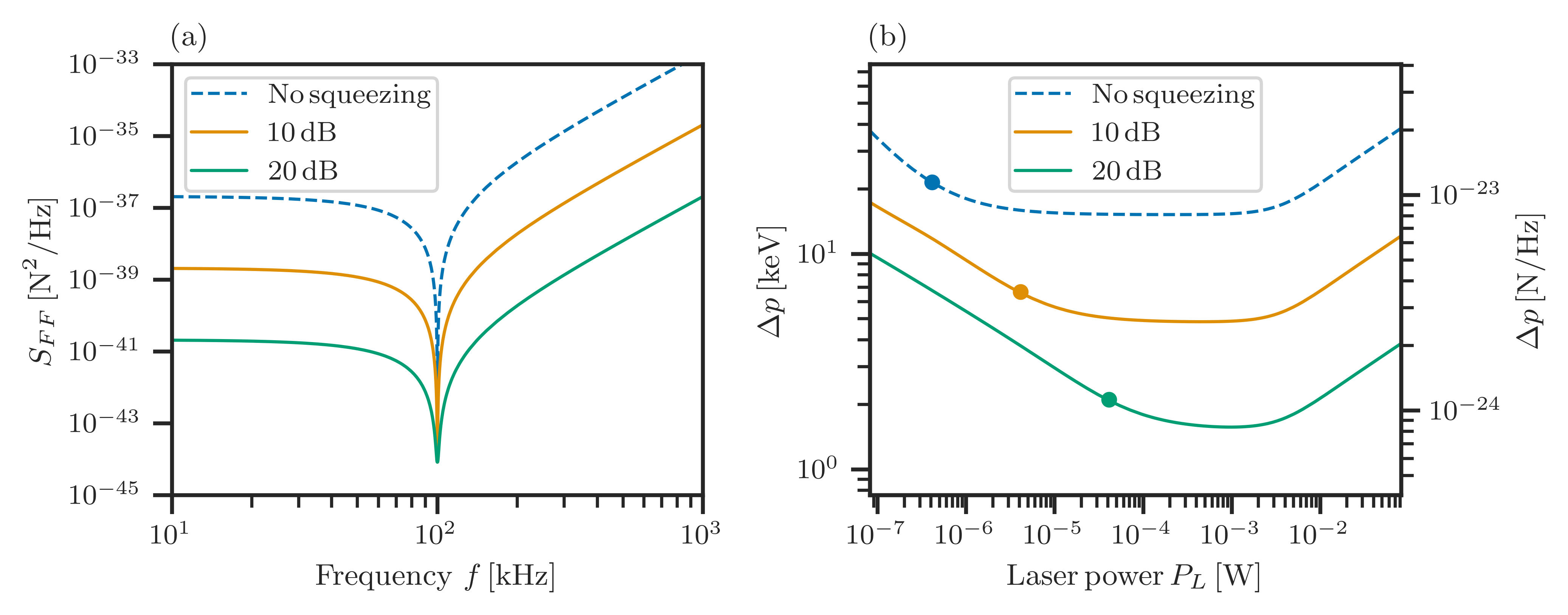}
\caption{(a) PSD of the suspended slab with the optimal frequency dependent squeezing. We show the PSDs without squeezing, and for 10 and 20 dB of input squeezing.  In each case, the power is set in order to achieve the SQL on-resonance, which corresponds to the powers indicated by circles in (b). (b) Momentum impulse detection threshold vs laser power. The circles indicate the parameters used to generate the PSDs of a). }
    \label{fig:FD_PSD}
\end{figure*}

We note here that even with the use of squeezed light, the noise on the mechanical resonance cannot go below the SQL. This is because $\mathrm{Re} \, \chi_m(\omega_m) = 0$, and so the cross-correlation term in Eq.~\eqref{eq:SYY-OM} vanishes.  The power required to reach the SQL on-resonance in the presence of squeezing is found as
\begin{align}
    &\frac{\partial S_{FF}(\omega_m; r, \theta_*)}{\partial g^2}\Big|_{g^2_*} = 0 \nonumber \\ &\implies g_*^2(\omega_m; r) = e^{2r} g_*^2(\omega_m; r=0).\label{eqn:squeezed_optimal_coupling}
\end{align}
We see that the optomechanical coupling to reach the SQL on-resonance is increased compared to the coupling required in the absence of squeezing.

If we can reduce the PSD in a broadband manner, the momentum threshold will improve. We see this graphically in Fig.~\ref{fig:FD_PSD}(b), where we observe that the numerically integrated momentum threshold decreases with more squeezing.

To gain an analytical understanding of the benefit of squeezing, we first observe in Fig.~\ref{fig:FD_PSD}(a) that the optimal $\Delta p$ again occurs for $g \gg g_*(\omega_m;r=0) = \sqrt{m \gamma \omega_m/2}$ in the slab case. We thus expand in the ratio $\tilde{g} \equiv g/g_*(\omega_m)$, suppressing the $r$-argument, to find
\begin{align}
    &S_{FF}(\nu) \approx \frac{e^{-2 r} m \omega_m^2}{2 Q} \times \tilde{g}^2 \nonumber \\
    &+ e^{-2 r} m\frac{e^{4 r} \nu^2+\frac{Q^2\left(\nu^2-\omega_m^2\right)^2}{\omega_m^2}}{2 Q} \times  \frac{1}{\tilde{g}^{2}} + m\omega_m \nu/Q +\ldots,
\label{eqn:couplingExpansionPSD}
\end{align}
where $Q = \omega_m/\gamma$.
This PSD is a quartic polynomial in $\nu$, as \eqref{eqn:coherentPSD} was, and so the integral we must perform is of the same functional form as \eqref{eq:threshold-sql}. Performing this integral, and expanding the result in $\sqrt{Q} \gg g/g_*(\omega_m) \gg 1$ , we find
\begin{equation}
\Delta p = e^{-r}\left(\frac{g^2+g^2_*(\omega_m)e^{2 r}}{g^2}\right)^{1 / 2}  \Delta p_\mathrm{SQL} + \mathcal{O}\left(\frac{\tilde{g}^2}{Q}\right)
\label{eqn:thresholdSqueezing}
\end{equation}Therefore, in the regime where $ \mathrm{e}^{r} \times g_*(\omega_m)\ll g \ll \sqrt{Q} \times g_*(\omega_m)$, we find that squeezing would simply decrease the impulse threshold by $\mathrm{e}^{-r}$. Moreover, the minimal power needed to achieve this exponential benefit to the momentum threshold increases with squeezing strength. For a low-$Q$ system, the regime does not exist, and one will not see the plateau feature as in Fig.~\ref{fig:FD_PSD}(b).

\begin{figure}[t]
\includegraphics[width = \linewidth]{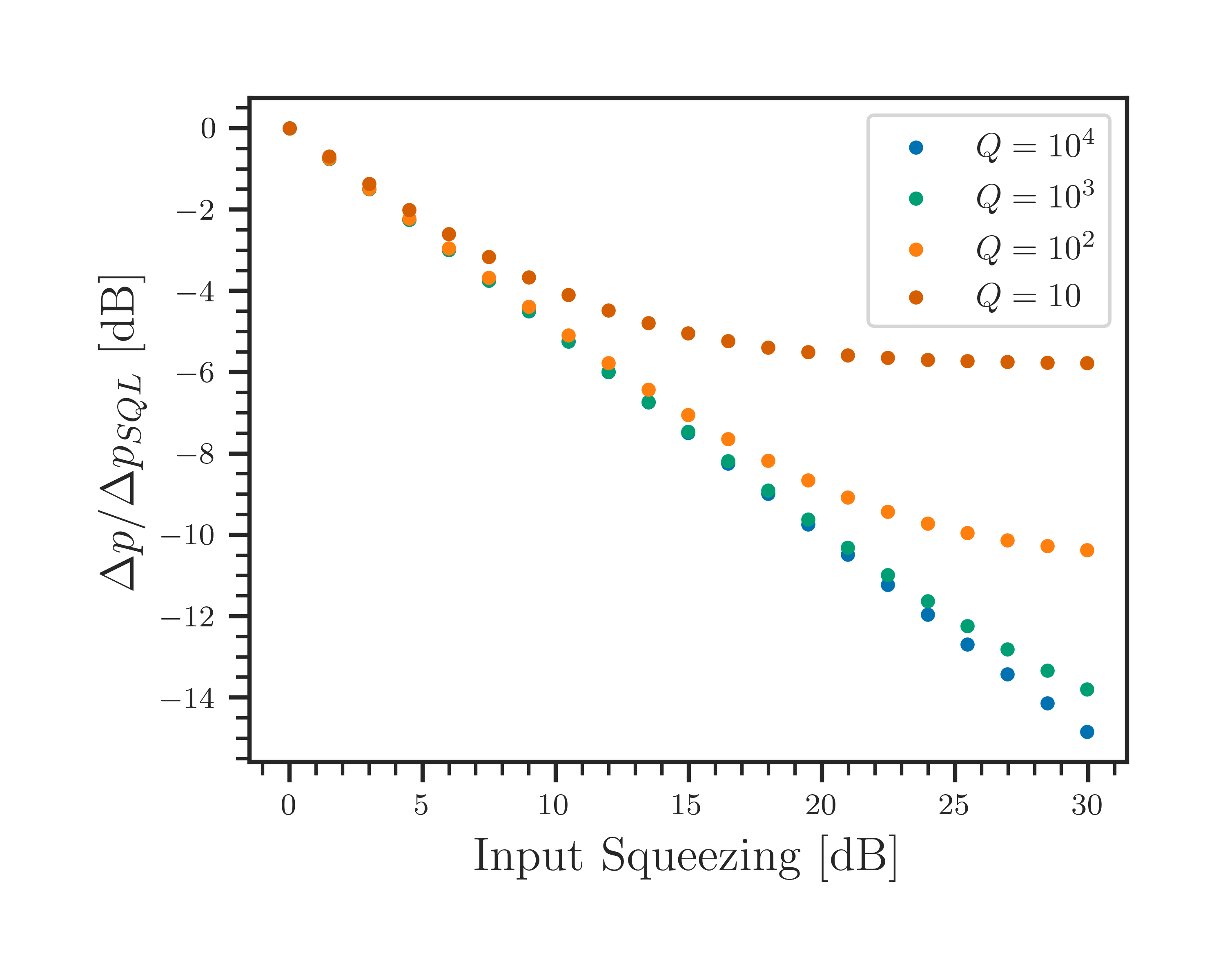}
    \caption{ The scaling of $\Delta p/\Delta p_\mathrm{SQL}$ with input squeezing [dB]. To convert between $r$ and dB, we note that $5r/\ln 10$ gives the amount of squeezing in dB. We take different values of $Q$, and we see that the plateau will be lower and appears at bigger $r$ for higher values of $Q$.  }
\label{fig:optimal_threshold}
\end{figure}

Another way to see this is to consider the optimally frequency-dependently squeezed PSD in the form of Eq. \eqref{eq:SFFopt}. In the $\tilde{g} = g/ g_*(\omega_m) \gg 1$ and $Q \gg e^{2r} \gg 1$ limit, we have
\begin{align}
    \frac{|\chi_{YX}^2 + \chi_{YY}^2 |}{|\chi_{YF}|^2} \approx  \frac{|\chi_{YX}|^2 + |\chi_{YY}^2 |}{|\chi_{YF}|^2} + \ldots,
\end{align}
and so
\begin{align}
    S_{FF}(\nu;r,\theta_*) \approx e^{-2r} S_{FF}(\nu; r=0),
\end{align}
up to terms of order $\sinh 2r/Q $ and $1/\tilde{g}^2$. Note that we have dropped the mass quantization noise contribution, as at large optomechanical couplings its contribution is everywhere subdominant to the measurement noise.
Since the PSD is everywhere exponentially decreased, the optimal momentum threshold similarly scales as
\begin{align}
    \Delta p(r) &\approx e^{-r}\Delta p(0) \nonumber \\ &\geq e^{-r} \Delta p_\mathrm{SQL}
    \label{eqn:FD_pTh}
\end{align}
We show this optimal behavior in Fig.~\ref{fig:optimal_threshold}. For small squeezing strengths, one sees indeed that the threshold decays as $\mathrm{e}^{-r}$. 

Once the squeezing strength becomes sufficiently large, Fig.~\ref{fig:optimal_threshold} shows there is a limit to the benefit that can be achieved with squeezing. As $r$ approaches \begin{equation}
    r_\mathrm{max} \approx \frac{1}{2}\mathrm{ln}\, Q
\end{equation}
 the large $Q$ expansion breaks down and the momentum threshold would gradually plateau. However, one can fully do the integration of \eqref{eqn:couplingExpansionPSD} and expand in large r instead of large Q, while keeping in mind that the optimal power that reaches the threshold minimum also scales with $e^r$. The result indicates that the minimum achievable momentum threshold is
\begin{align}
    \Delta p _\mathrm{min} \approx \frac{\Delta p_\mathrm{SQL}}{\sqrt{Q}}.
    \label{eqn:fund_limit}
\end{align}
 This sets an intrinsic limit to the sensitivity of these devices using squeezed light strategies that depends on the quality of the mechanical oscillator. In this limit, the analytics leading to \eqref{eqn:couplingExpansionPSD} has broken down, as the condition $\sinh \, 2r /Q \ll 1$ is no longer satisfied, leading to $\mathcal{O}(1)$ corrections that give rise to the plateau seen in Fig.~\ref{fig:optimal_threshold}. 
 
 For many physical systems of interest, $e^{2r} \ll Q$, and so one still expects to reduce the measurement noise through the use of higher squeezing. The quality factor of free-space, optomechanical set-ups ranges from as low as $7$~\cite{wang2024Mechanical} to as high as $10^{7}$~\cite{hofer2023high}, while the generation of squeezing is limited to 15 dB~\cite{vahlbruch2008observation, vahlbruch2016detection}. The upper limit of high-Q mechanics is well within the $e^{2r} \ll Q$ limit, while the lower-Q systems (where $e^{2r} \sim Q$) is in fact limited by additional feedback damping, which must be reduced to take full advantage of squeezing. Ultimately, this result implies that any coupling to the environment that leads to damping precludes a perfect measurement of the system.

The result of Eq. \eqref{eqn:fund_limit} is similar to existing results in the literature~\cite{jaekel1990quantum, braginsky1995quantum,danilishin2012quantum}, in that it is the dissipative part $\gamma$ of the system response that limits the fundamental floor of the measurement noise. In interferometric systems, for instance, one can show that noise floor achievable in measuring a free mass with frequency-dependent squeezing is set by the cavity damping rate~\cite{danilishin2012quantum}, and we here have extended this to the free-space measurement of a damped, harmonic oscillator.         Furthermore, in the context of GW detection, it is known that the dissipative part of the mechanical susceptibility also limits the efficacy of FD squeezing~\cite{jaekel1990quantum}.

        Our result may be compared with the fundamental lower bound on the force PSD set by Heisenberg uncertainty of twice the quantization noise $S_{FF} \geq 2 m \gamma \nu$~\cite{braginsky1995quantum}. Both this result and our Eq. \eqref{eqn:fund_limit} are limited by the mechanical dissipation, although we see that frequency-dependent squeezing does not obtain this limit, even for arbitrarily large squeezing.
Moreover, this may be considered as a specific realization of the limits of waveform estimation found in ~\cite{tsang2011fundamental, miao2017towards}, when applied to our problem of inferring a delta-function force from a continuous sequence of position measurements.

\section{Losses}

In this section, we investigate how losses affect the achievable sensitivity. In realistic systems, losses arise due to imperfect photodetection efficiencies. These may arise both from the finite efficiency intrinsic to the detectors, or, in the case of three-dimensional systems, from the fact that not all photons are scattered into a photodetector. We model losses in the usual way by a fictitious beamsplitter~\cite{fearn1990linear, leonhardt2003quantum}
\begin{align}
    Y^{\mathrm{out}}_\eta (\nu) &= \sqrt{\eta} Y^\mathrm{out}(\nu) + \sqrt{1- \eta} \tilde{Y}^\mathrm{in}(\nu) \nonumber \\
    &\supset \sqrt{\eta} \, \chi_{YF} F^\mathrm{in},
    \label{eqn:lossyOutput}
\end{align}
where $Y^{\mathrm{out}}_\eta$ is the measured phase quadrature that depends on a detection efficiency parameter $\eta \in [0,1]$, and $\tilde{Y}^\mathrm{in}(\nu)$ is the added vacuum noise that satisfies $S_{\tilde{Y}\tilde{Y}} = 1/2$. In the second line of \eqref{eqn:lossyOutput}, we have isolated the dependence of the phase on the input force, from which we see that in the presence of losses an unbiased estimator of the force is obtained by
\begin{align}
    F_{\mathrm{E},\eta} = \eta^{-1/2} \frac{Y^\mathrm{out}}{\chi_{YF}},
\end{align}
which is simply a rescaling of the lossless estimator.
The resulting PSD of this force estimator, including losses, is 
\begin{equation}
S_{FF}(\nu; \eta)=S_{FF}(\nu; \eta=1)+\frac{1-\eta}{\eta |\chi_{YF}|^2}S_{\tilde{Y}\tilde{Y}}, 
\label{eqn:lossyPSD}
\end{equation}
which is the sum of the lossless PSD $S_{FF}(\nu; \eta=1)$ and an extra loss term.

\begin{figure*}[ht!]
    \centering
    \includegraphics[width = \linewidth]{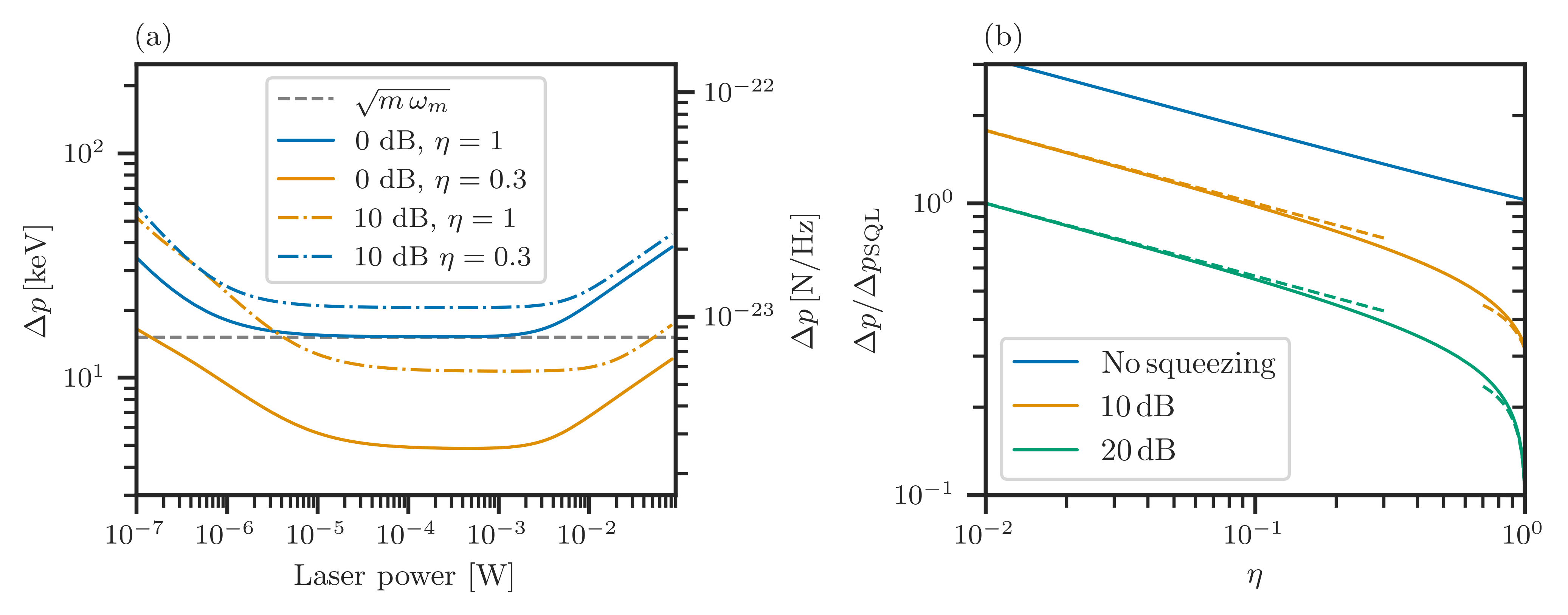}
\caption{(a) The impulse threshold for a dielectric slab as a function of the laser power, where we include the effects of loss. (b) The optimal momentum threshold $\Delta p$ relative to $\Delta p_\mathrm{SQL} = \sqrt{m \omega_m}$ for a system with detection efficiency $\eta$, evaluated numerically (solid) and analytically (dashed). The analytic solution corresponds to the small-$\eta$ asymptotics \eqref{eqn:lossyPth} for $\eta \leq 0.3$, and the small-$(1-\eta)$ asymptotics \eqref{eqn:largeEtaExpansion} for $1-\eta \leq 0.3$. We can see from the spacing on both sides of the y-axis that the effect of squeezing is different between the lossy and lossless regimes: in the former case, the $r$-scaling is $\Delta p \sim e^{-r/2}$, while in the latter it is $\Delta p \sim e^{-r}$. }
    \label{fig:slab-threshold-minima}
\end{figure*}
\subsection{Effect of losses on vacuum noise}

We first consider the consequence of the extra noise in a lossy, $\eta \neq 1$ system in the absence of squeezing. In this case, all the input PSDs satisfy their vacuum values with $S_{XX} = S_{YY} = S_{\tilde{Y} \tilde{Y}}$, and so \eqref{eqn:lossyPSD} is given by
\begin{align}
    S_{FF}(\nu; \eta) = |\chi_{YF}|^{-2} \left( \frac{S_{YY}}{\eta} + |\chi_{YX}|^2 S_{XX} \right),
\label{eqn:lossyCoherentPSD}
\end{align}
which we see acts purely as a rescaling of the shot noise. 

How does this affect the optimal momentum threshold $\Delta p$ that may be achieved? In terms of the optomechanical coupling $g$, the shot-noise scales as $g^{-2}$ while the back-action scales as $g^2$. From these two facts, we observe the PSD \eqref{eqn:lossyCoherentPSD} may be written in terms of the PSD with a rescaled coupling
\begin{align}
    S_{FF}(\nu;\eta,g) = \eta^{-1/2} S_{FF}(\nu; \eta = 1,\eta^{-1/4} \, g ),
\end{align}
which is proportional to the lossless PSD $S_{FF}(\nu;\eta = 1,\eta^{-1/4} \, g)$ with a coupling that is $\eta^{-1/4}$ times larger than $g$. Therefore given a coupling $g_*$ that optimizes the lossless momentum sensitivity $\Delta p(\eta=1)$, the use of an increased coupling $g = g_* \eta^{-1/4}$ gives the optimal sensitivity
\begin{equation}
    \Delta p(\eta) = \eta^{-1/4} \Delta p_\mathrm{SQL},
    \label{eqn:lossypTh}
\end{equation}
which exhibits a relatively soft dependence on the loss parameter $\eta$ at the cost of a modest increase in laser power.

\subsection{Effect of losses on squeezed noise}

We now consider the interplay of losses and squeezed input light. We first note that the second term in \eqref{eqn:lossyPSD} is proportional to a constant $S_{\tilde{Y}\tilde{Y}}$ that is independent of the input light. Since it does not depend on the squeezing angle $\theta$, the optimal squeezing angle $\theta_*(\nu)$ is unaffected by losses. In this case, 
\begin{align}
    S_{FF}(\nu;r,\theta_*;\eta) = S_{FF}(\nu;r,\theta_*;1) + \frac{1-\eta}{2\eta |\chi_{YF}|^2},
\end{align}
where $S_{FF}(\nu;r,\theta_*;1)$ is given by ~\eqref{eq:SFFopt}. 

As the optimal squeezing angle remains the same, the arguments of Sec.~\ref{sec:FD_squeezing} allow us to make the large $Q$ and large $g/g_*(\omega_m;r=0)$ approximation 
\begin{align}
    S_{FF}(\nu;r,\eta) &\approx e^{-2r} S^0_{FF} + \frac{m \nu^2}{Q c^2} \sinh 2r +  \frac{1-\eta}{2\eta |\chi_{YF}|^2} + m \gamma \nu \nonumber \\
    &\approx e^{-2r} S^0_{FF} + \frac{1-\eta}{2\eta |\chi_{YF}|^2},
    \label{eqn:lossyPSD2}
\end{align}
where in the first line we have made use of \eqref{eqn:coherentPSD} and \eqref{eqn:couplingExpansionPSD}, and the second line follows in the limit that $ g \gg e^r g_*(\omega_m;0)$, which we know is necessary to optimize the momentum sensitivity from \eqref{eqn:thresholdSqueezing}. This shows us that, at best, squeezing will exponentially decrease the coherent-state measurement noise $S_{FF}^0$, while the shot-noise from inefficiencies is unaffected by squeezing.

In general, the impulse threshold arising from \eqref{eqn:lossyPSD} is difficult to evaluate analytically. However, for small detection efficiencies $\eta \ll 1$, the shot-noise contribution to \eqref{eqn:lossyPSD2} is dominated by the effect of losses, and so the force PSD may be expressed as
\begin{align}
    S_{FF}(\nu;r,\eta) & \approx \frac{1}{2|\chi_{YF}|^2} \Big[ e^{-2r}\left( 1 + |\chi_{YX}|^2 \right) + \frac{1-\eta}{\eta} \Big] \nonumber \\ &\approx \frac{\eta^{-1/2}e^{-r}}{2|\chi_{YF}|^2} \Big[ \frac{1}{e^{-r} \eta^{1/2}} + e^{-r}\eta^{1/2} |\chi_{YX}|^2 \Big], 
    \label{eqn:lossyPSD3}
\end{align}
where in the second line we have dropped the sub-leading contribution to the shot-noise. Writing the PSD in this way, we see by comparison with \eqref{eqn:coherentPSD} that \eqref{eqn:lossyPSD3} is $\eta^{-1/2}e^{-r}$ times the lossless, coherent-state PSD, with a rescaling of $g \to e^{r/2} \eta^{-1/4} g$. Therefore, it follows by the arguments of Sec.~\ref{sec:coherentState} that in this regime of small efficiencies $\eta$, the threshold scales as
\begin{align}
    \Delta p_\mathrm{thresh} (r,\eta) \approx \eta^{-1/4} e^{-r/2} \Delta p_\mathrm{SQL},
    \label{eqn:lossyPth}
\end{align}
provided $ g \gg e^{r/2} \eta^{-1/4} g_*(\omega_m)$. This demonstrates that the scaling with efficiency $\eta$ has the same $-1/4$ index as the coherent-state strategy in \eqref{eqn:lossypTh} at small efficiencies, which is reassuring. Furthermore, the noise can still be made arbitrarily low: the shot noise is dominated by the losses, which may be mitigated by increasing the laser power, which in turn may be mitigated by squeezing the back-action. The losses have dramatically affected the scaling with the squeezing parameter, however, as the minimal noise now scales as $e^{-r/2}$ rather than $e^{-r}$.

One can also find an analytic result for almost lossless systems, i.e., for $1-\eta \ll 1$. In this limit, \eqref{eqn:lossyPSD}
is almost that of a lossless system, but there is a small increase to the shot-noise. We may then perform the integral over the inverse of this quartic PSD as in \eqref{eq:threshold-sql} to find 
\begin{align}
    \Delta p_\mathrm{thresh}(r,\eta) \approx \big[ 1+ (1-\eta)e^{2r}\big]^{1/4} e^{-r} \Delta p_\mathrm{SQL},
    \label{eqn:largeEtaExpansion}
\end{align}
which we see still has the same $e^{-r}$ scaling as the lossless case as long as $(1-\eta)e^{2r} \ll 1$. Once the degree of squeezing becomes large enough that $(1-\eta)e^{2r} \gg 1$, even while $1 - \eta \ll 1$, the threshold \eqref{eqn:largeEtaExpansion} still tends to a scaling $\Delta p \sim e^{-r/2}$, demonstrating the ultimate sensitivity of squeezing to losses. 

In Fig.~\ref{fig:slab-threshold-minima}(b), we plot the threshold, evaluated numerically with an optimized coupling $g$ and compared to to the analytic scalings. Both the analytic asymptotics of \eqref{eqn:lossyPth} and \eqref{eqn:largeEtaExpansion} are good approximations in their regimes of validity. In particular, we note observation of the transition to the $\Delta p \sim e^{-r/2}$ scaling regime at small $\eta$, where the benefits of squeezing persist but are lessened.

\section{Outlook}

In this work, we studied quantum-enhanced strategies for measuring impulsive signals of harmonically suspended dielectrics in detail. Once all other technical and environmental noise sources are sufficiently mitigated, one is limited by the uncertainty intrinsic to the light used to make the measurement itself. We have shown that this noise may be reduced below the SQL by the use of squeezed light, and that frequency-dependent squeezing is required to achieve optimal scaling with $r$ due to the broadband nature of the signal. We have shown that this conclusion is robust to the presence of optical losses.

Our central result is that with a fixed amount of frequency-dependent squeezing (say $n$ dB reduction at all frequencies in the input noise PSD), the achievable momentum threshold $\Delta p$ is reduced by an equivalent amount compared to the value at the SQL. This arises because for a squeezing parameter $r$, frequency-dependent squeezing can suppress the off-resonant sections of the PSD by $e^{-2r}$, while, on resonance, the PSD is always bounded by the SQL value $S_{FF}(\omega_m) \geq 2m\gamma\omega_m$. 

We also found a fundamental limit to this effect set by the mechanical quality factor of the detector: for a low Q oscillator, or for extremely large amounts of squeezing, there is a limit on the benefit squeezing on the impulse threshold, as seen in Fig.~\ref{fig:optimal_threshold}. The minimum achievable momentum threshold is given by $\Delta p_{\rm min} = \Delta p_{\rm SQL}/\sqrt{Q}$, although for a high-$Q$ oscillator this threshold is usually out of reach with realistic levels of squeezing.

Finally, we quantified the effects of realistic optical loss on the advantages generated by squeezed light. In the limit of reasonably low loss (photodetection efficiency $\eta \approx 1$), the ideal scaling predicted in Eq.~\eqref{dp-expected} can be achieved up to small errors, quantified in Eq.~\eqref{eqn:largeEtaExpansion}. In the limit of large loss (efficiency $\eta \approx 0$) or large squeezing ($e^{2r}(1-\eta) \gg 1)$, however, the scaling of $\Delta p$ was exponential in $-r/2$, rather than $-r$. Interestingly, this implies that one can still achieve sub-SQL thresholds even with substantial optical loss, but with worse scaling as a function of squeezing strength. 

Moving forward, a few clear directions emerge based on the above results. Extending the analysis from the simple 1-dimensional slab models above to a 3-dimensional scattering nanosphere would imply a minimal amount of photon loss, because in general it is impossible to do a full $4\pi$ photodetection. Thus, the benefit of squeezing will be limited. This motivates an alternative approach to sub-SQL sensing based on directly squeezing the mechanical motion itself~ \cite{rossi2024quantumdelocalizationlevitatednanoparticle,PhysRevA.46.6091}, which might not suffer so severely from these drawbacks. Pulsed measurements~\cite{vanner2011pulsed}, which would induce less decoherence through photon recoil, may also be beneficial. We leave this for future work.

\section*{Acknowledgements}

We thank David Moore for useful dicussions. We thank the Kavli Institute for Theoretical Physics (supported in part by the National Science Foundation under Grants No. NSF PHY-1748958 and PHY-2309135) for hospitality while part of this work was completed. Our work at LBL is supported by the U.S. DOE, Office of High Energy Physics, under Contract No. DEAC02-05CH11231, the Quantum Information Science Enabled Discovery (QuantISED) for High Energy Physics grant KA2401032, the Quantum Horizons: QIS Research and Innovation for Nuclear Science Award DE-SC0023672, and an Office of Science Graduate Student Research (SCGSR) fellowship (administered by the Oak Ridge Institute for Science and Education for the DOE under contract number DE‐SC0014664).

\newpage
\bibliographystyle{utphys-dan} 
\bibliography{main}

\newpage

\appendix

\section{Conventions}

We take the following Fourier transform conventions:
\begin{align}
    \tilde{f}(\nu) &= \int dt e^{i \nu t} f(t), \\
    f(t)& = \int \frac{d\nu}{2\pi}e^{-i\nu t} \tilde{f}(\nu),
\end{align}
where all integrals are along the whole real line, unless otherwise stated. PSDs are defined as
\begin{align}
    2 \pi \delta(\nu + \nu') S_{\mathcal{O}\mathcal{O}}(\nu) = \langle \mathcal{O}(\nu) \mathcal{O}(\nu')\rangle.
\end{align}

\section{Fabry-P\'erot cavity details}
\label{app:FP-details}
The impulse sensing results with cavity opto-mechanical sensors are analytically simpler than in free space because the cavity selects essentially a single mode of the input light field, thus collapsing many sums that would otherwise be present throughout. However, such systems still contains much of the essential physics needed to understand our later results, thus we begin by analyzing a canonical cavity opto-mechanical sensor: a Fabry-P\'erot cavity.  This system is analyzed carefully in many places (see for example Ch.~2 of Ref.~\cite{bowen2016quantum} or Sec. IV of Ref.~\cite{beckey2023quantum}), so we provide only the essential features here.

As depicted in Fig.~\ref{fig:FP-and-slab}, the Fabry-P\'erot cavity is characterized by a mechanical frequency $\omega_c$, a cavity loss rate $\kappa$, and bare length $L$. We will introduce standard bosonic creation and annihilation operators $b^{\dagger},b$ for the mechanics and $a^{\dagger},a$ for the cavity mode, respectively. The equations of motion will be expressed in terms of  the amplitude and phase quadratures of the light given as
\begin{align}
    X=\frac{a+a^{\dagger}}{\sqrt{2}} \quad \text{and} \quad Y=\frac{a-a^{\dagger}}{i\sqrt{2}},
\end{align}
and the mechanical position and momentum given as
\begin{align}
    x=x_0(b+b^{\dagger}) \quad \text{and} \quad p=p_0(b-b^{\dagger}),
\end{align}
where $x_0 = 1/\sqrt{2 m \omega_m}$ and $p_0 = \sqrt{m \omega_m/2}$ are the widths of the ground state of the harmonic oscillator in phase space.

The opto-mechanical coupling, which quantifies how quickly the cavity frequency changes with cavity length, is given as 
\begin{align}
    g_0 = \frac{\omega_c}{L}.
\end{align}
Typically this coupling is enhanced by driving the interaction with a laser (of power $P_L$) containing an average photon number given by $|\alpha|^2 = P_L/\omega_c \kappa$, leading to an enhanced coupling given by $g_c = |\alpha|g_0$. Linearizing around this strong background drive, the Hamiltonian describing this system becomes

\begin{align}
\label{eq:H-OM}
H_{\rm det} = \frac{p^2}{2m} + \frac{1}{2} m \omega_m^2 x^2 + \omega_c a^\dag a + \sqrt{2} g_c x X.
\end{align}
We see that the linearized opto-mechanical Hamiltonian simply describes position-position coupled oscillators, leading to the Heisenberg equations of motion given as
\begin{align}
\begin{split}
    \dot{X} &= \Delta Y - \frac{\kappa}{2} X + \sqrt{\kappa} X^{\rm in},\\
    \dot{Y}  &= - \Delta X - \frac{\kappa}{2} Y + \sqrt{\kappa} Y^{\rm in}  - \sqrt{2}g x,\\
    \dot{x} &= \frac{p}{m},\\
    \dot{p} &= -m\omega_m^2 x - \gamma p + F^{\rm in} - \sqrt{2}g_cX.
\end{split}
\label{eq:EOM-OM}
\end{align}

Together with the input-ouput relations given as 
\begin{align}
\begin{split}
\label{eq:X-I/O}
     X^{\rm out} &= X^{\rm in} - \sqrt{\kappa} X,\\
    Y^{\rm out} &= Y^{\rm in} - \sqrt{\kappa} Y. 
\end{split}
\end{align}
We can solve the equations in the frequency domain to obtain, for example, the output phase quadrature in terms of the input fields as
\begin{align}\label{eq:Y-out-linear-system}
Y^\mathrm{out} &= \chi_{YY} Y^\mathrm{in} + \chi_{YX}  X^\mathrm{in} + \chi_{YF} F^\mathrm{in},
\end{align}
where we have defined the susceptibilities relating input to output fields as
\begin{align} 
\begin{split}
\label{eq:transfer-OM}
    \chi_{YY}(\nu) &= 1+\kappa \chi_c(\nu) = e^{i\phi_c(\nu)}\\
    \chi_{YX}(\nu) &= - 2 \kappa g_c^2 \chi_c^2(\nu) \chi_m(\nu),\\
    \chi_{YF}(\nu) &= -(2 \kappa g_c^2)^{1/2} \chi_c(\nu) \chi_m (\nu),
\end{split}
\end{align}
which themselves depend on the cavity and mechanical susceptibilities
\begin{align}
\begin{split}
\label{susc-OM}
    \chi_c (\nu) &= \frac{1}{i \nu-\kappa/2},\\
    \chi_m (\nu) &= \frac{1}{m(\omega_m^2 - \nu^2 - i \gamma \nu)}.
\end{split}
\end{align}
In practice, this quadrature will be accessed using homodyne interferometry, which leads to a data stream from which we can construct an estimator for the force on the mirror via
\begin{align}
\label{eq:FE-OM}
F_E(\nu) = \frac{Y^\mathrm{out}(\nu)}{\chi_{YF}(\nu)}.
\end{align}
Finally, to obtain the momentum threshold, we must compute the PSD for this estimator as
\begin{align}
\begin{split}
S^{\rm out}_{YY} = & |\chi_{YY}|^2 S^{\rm in}_{YY} + |\chi_{YX}|^2 S^{\rm in}_{XX} + |\chi_{YF}|^2 S^{\rm in}_{FF} \\
& + \chi_{YX} \chi^*_{YY} S^{\rm in}_{YX} + \chi_{YY} \chi^*_{YX} S^{\rm in}_{XY},
\end{split}
\end{align}
leading to 
\begin{align}\label{eq:SFF-OM}
    S_{FF}(\nu) = \frac{S_{YY}^{\rm out} (\nu)}{|\chi_{YF}(\nu)|^2}.
\end{align}
We see from inspection of Eq.~\eqref{eq:SYY-OM} that the noise is comprised of shot-noise, back-action, thermal, and a cross-correlation between shot-noise and back-action. For the time being, we neglect the thermal noise contribution, $S_{FF}^{\rm in})$, and focus only on quantum noise. In the absence of squeezing, $S_{YX}^{\rm in} = S_{YX}^{\rm in}=0$ and the noise is simply a trade-off between shot noise and back-action. In this setting, it is easy to see that the shot noise and back action can be minimized at a particular frequency, $\omega_*$, to yield the SQL. 

With squeezing, however, the cross-terms in the PSD become
\begin{align}
\begin{split}
    S^{\rm in}_{XX}&= \frac{1}{2}[\cosh{2r} -\cos{\theta}\sinh{2r}]  ,\\
    S^{\rm in}_{XY}& = S^{\rm in}_{YX} = -  \frac{1}{2}[ \sinh{2r} \sin{\theta} ] ,\\
    S^{\rm in}_{YY}&= \frac{1}{2}[ \cosh{2r} +\cos{\theta}\sinh{2r}].
\end{split}
\label{eqn:SXX_squeezed}
\end{align}

\begin{figure*}
    \centering
    \includegraphics[width=0.5\linewidth]{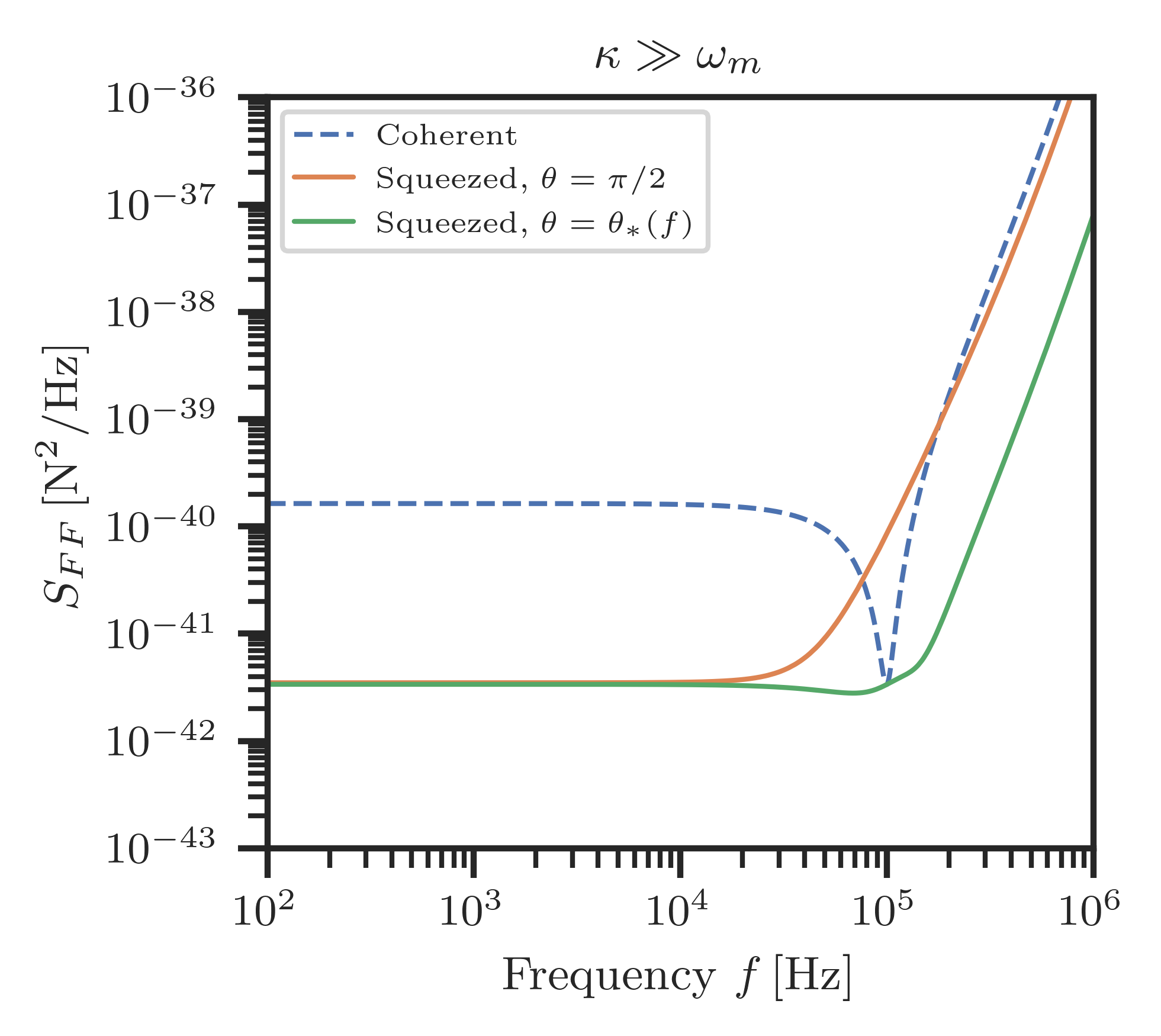}\hfill
    \includegraphics[width=0.5\linewidth]{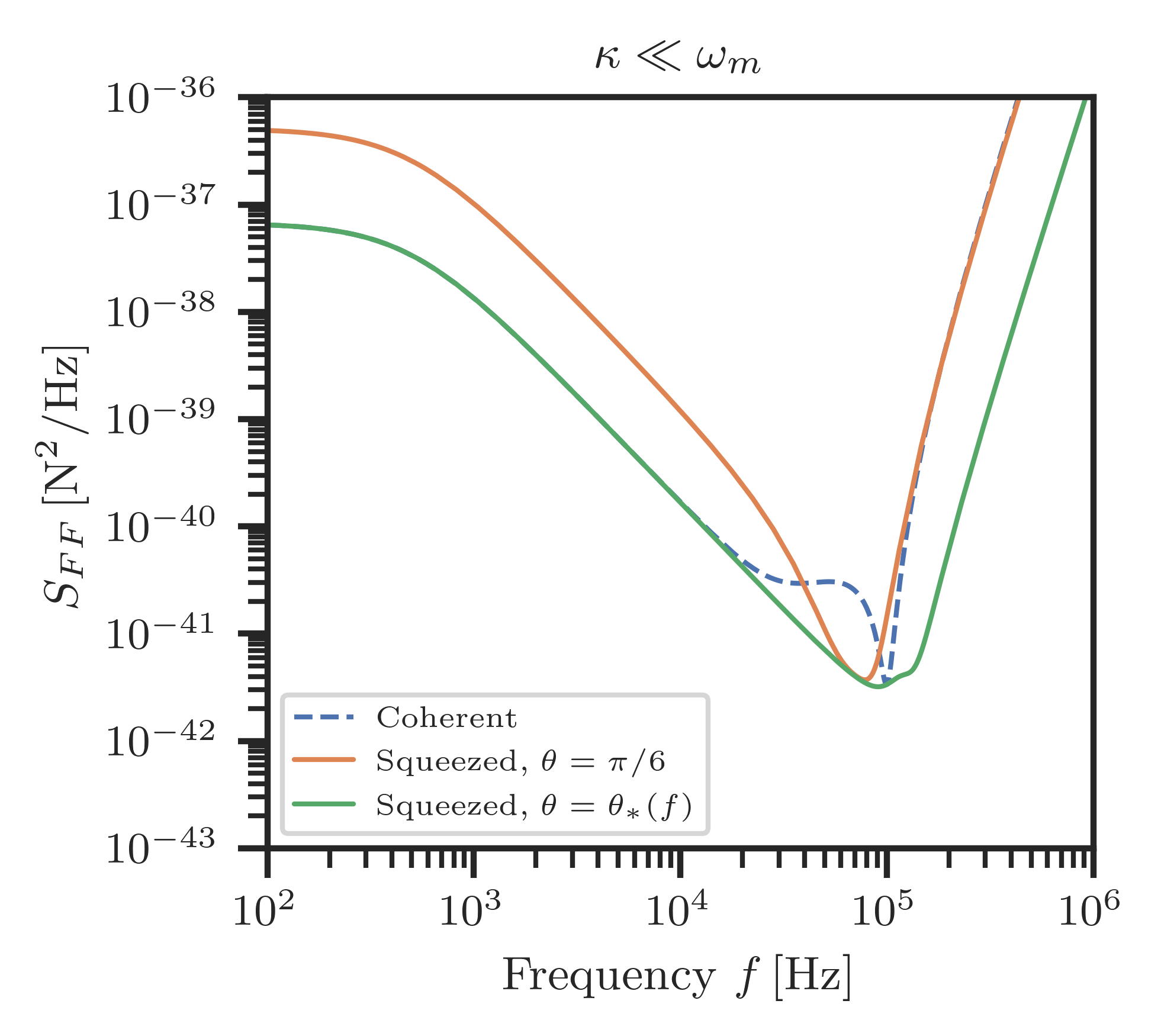}
\caption{The PSDs of the force estimator for an optomechanical cavity in the bad cavity (left) and good cavity (right) limits.}
    \label{fig:cavityPSDs}
\end{figure*}

Importantly, unlike the vacuum, the squeezed vacuum state has cross-correlations $S_{XY} \neq 0$ between the two quadratures. Moreover, $S_{XY}$ can be negative, unlike $S_{XX}$ and $S_{YY}$. Furthermore, note that, in general both $\theta$ and $r$ will have some dependence on frequency. In this work, we will explicitly denote the frequency dependence in the squeezing angle, $\theta(\nu)$, and will assume that $r$ is frequency-independent over the relevant range of frequencies. We plot in Fig.~\ref{fig:cavityPSDs} the shape of the cavity PSDs for a good and bad cavity for a variety of squeezing strategies in order to illustrate the difference between the two cavity limits. 

While it has been known for nearly half a century that squeezing can lower a force PSD~\cite{caves1981quantum}, the benefit of various squeezing strategies to impulse sensing remains far less well understood. Thus, we now take the time to derive derive our figure of merit, the momentum threshold, which will tell us the minimal resolvable momentum kick that can be detected by our sensor.

\section{Fluctuation-dissipation theorem}
\label{app:FD_theorem}

In this appendix, we discuss the contribution to the power spectral density arising from the quantum mechanical nature of the mechanical element and its damping, which must give rise to added noise by the fluctuation-dissipation theorem~\cite{kubofluctuationdissipation1966}. 

To treat the intrinsic noise due to test mass quantization, we follow~\cite{Whittle:2022gqe} by starting with the fluctuation dissipation theorem. We may model this by standard input-output theory as a coupling of the position $x$ to a stochastic force $F_x$ given by the bath-operator interaction
\begin{align}
    H_{\mathrm{b}x} = - x F_x.
    \label{eqn:bathInteraction}
\end{align}
The fluctuation-dissipation theorem states that within the linear response approximation, symmetrised PSDs $S_{\mathcal{O}x}$ between some observable $\mathcal{O}$ and the operator $x$ obey \cite{kubofluctuationdissipation1966}
\begin{align}
    S^\mathrm{bath}_{\mathcal{O} x}(\nu) = \frac{\mathrm{Im} \, \chi_{\mathcal{O}x}(\nu)}{\tanh{\left(\frac{\nu}{2T}\right)}},
    \label{eqn:fluctDiss}
\end{align}
where $\chi_{\mathcal{O}x}(\nu)$ is the susceptibility describing the modified behaviour of the observable $\mathcal{O}$ due to the bath:
\begin{align}
    \delta \mathcal{O}(\nu) = \chi_{\mathcal{O}x}(\nu) F_x(\nu).
\end{align} 
For the particular case of the position PSD $S_{xx}$, the susceptibility is simply the mechanical susceptibility $\chi_{xx} = \chi_m$, and so
\begin{align}
    S^\mathrm{bath}_{xx} = \frac{\mathrm{Im} \, \chi_m}{\tanh{\left(\frac{\nu}{2T}\right)}}.
\end{align}

The quantization contribution to the force PSD arising from the dissipative interaction with the bath is given by the zero-temperature limit of
\begin{align}
    S^\mathrm{bath}_{FF} = \frac{S^\mathrm{bath}_{xx}}{|\chi_m|^2} =  \coth \frac{\omega}{2 T} \frac{\mathrm{Im} \, \chi_m}{|\chi_m|^2},
\end{align}
namely
\begin{align}
    S^\mathrm{QN}_{FF} =  \frac{\mathrm{Im} \, \chi_m}{|\chi_m|^2} = m \gamma \nu.
    \label{eqn:QN}
\end{align}
Equation \eqref{eqn:QN} represents the minimal added noise due to the dissipative coupling of the mechanical element of the bath allowed by quantum mechanics, within the weakly coupled regime where linear response is appropriate.

\section{Two-port slab PSD details} \label{app:slab-details}

\subsection{Impulse sensing with a harmonically trapped dielectric slab}

We will consider a planar dielectric slab suspended harmonically in free space. In practice, the trapping potential could be formed by a mechanical suspension system, as depicted in Fig.~\ref{fig:FP-and-slab}(b), or it could be supplied by optical tweezers, for example. This physical system captures some of the essential features of the more complicated case of free-space nanospheres trapped in optical tweezers~\cite{romero2011optically,Maurer:2022nvj,gonzalez-ballestero2023Suppressing,magrini2021Realtime}, which have been proposed to be used as extremely sensitive impulse detectors~\cite{Carney:2022pku}. Our treatment here is based on that in \cite{beckey2023quantum}, although with quite a few important refinements.

Consider a slab of mass $m$, electric susceptibility $\chi_\mathrm{e}$, mechanical resonance frequency of $\omega_m$, thickness $\ell$, and infinite planar extent, as shown in Fig.~\ref{fig:FP-and-slab}(b). Assuming that the slab is linearly polarizable, i.e. $\mb{P}(\mb{r}) = \chi_\mathrm{e} \mb{E}(\mb{r})$, the light-slab interaction Hamiltonian is given by
\begin{equation}
V_{\text {int }}(x, t)=-\frac{1}{2}\chi_\mathrm{e} \int_{\text {slab }} d^3 \mathbf{r}|\mathbf{E}(\mathbf{r}, t)|^2,
\label{eq:V-int}
\end{equation}
    where the susceptibility $\chi_\mathrm{e}$ may be expressed in terms of the dielectric constant $\epsilon_r$ as $\chi_\mathrm{e} = \epsilon_r-1$.

We wish to express this interaction Hamiltonian in terms of an interaction between the modes of light and the position $x$ of the slab. An essential difference here compared to the cavity considered above is that in free space one must consider the dynamics of many modes. We will decompose these modes into plane-waves, since the planar slab symmetry ensures that these are eigenmodes of the system, in contrast to the more complicated case of the spherical nano-particles~\cite{Maurer:2022nvj}. Moreover, because we may want to monitor both the left- and right-moving modes separately, we will explicitly separate the light into right-moving modes and left-moving modes. We now decompose the free-field in terms of its continuum modes as
\begin{align}
\begin{split}
E(\mathbf{r}) & =E_R(\mathbf{r})+E_L(\mathbf{r}), \\
E_R(\mathbf{r}) & =\frac{i}{(2 \pi)^3} \int d^3 \mathbf{k}\sqrt{\omega_{\mathbf{k}}}  \left[e^{-i\left(k r_x+\mathbf{k}_{\perp} \cdot \mathbf{r}_{\perp}\right)} a_{k, \mathbf{k}_{\perp}}-\rm{c.c.}\right], \\
E_L(\mathbf{r}) & = \frac{i}{(2 \pi)^3} \int d^3 \mathbf{k}\sqrt{\omega_{\mathbf{k}}}  \left[e^{i\left(k r_x-\mathbf{k}_{\perp} \cdot \mathbf{r}_{\perp}\right)} b_{k, \mathbf{k}_{\perp}}-\rm{c.c.}\right].
\end{split}
\label{eq:lrdef}
\end{align}
where  
$\omega_{\mathbf{k}}=\sqrt{k^2+\mathbf{k}_{\perp}^2}$, and $k$ ($\mb{k}_\perp$) are parallel (perpendicular) to the slab's normal vector. We take the commutation relations $\left[a_\mathbf{k}, a_{\mathbf{k}^{\prime}}^{\dagger}\right]=\delta\left(\mathbf{k}-\mathbf{k}^{\prime}\right)$. Here we have ignored the polarization of light, as we take the background field around which we will expand to be linearly polarized. 

Now that we have expressions for the light modes, we may see how these interact with the slab. Focusing on small displacements of $x\ll \lambda$, we can expand Eq.~\eqref{eq:V-int} as
\begin{align}
\label{app-taylor}
V(x) = V(0) + x \partial_x V(0) + \mathcal{O}(x^2).
\end{align}
Ignoring the position-independent renormalization, the linear term in $x$ is the dominant interaction.\footnote{There are also terms proportional to $x^2$, which gives a harmonic trapping potential. Here for simplicity we have modelled the slab as externally suspended with fixed mechanical frequency $\omega_m$, part of which comes from this laser potential.} This interaction can be written as
\begin{align}
\begin{split}
& x \partial_x V(0)\\
&=-\frac{\chi_\mathrm{e} x}{2} \int d^2 \mathbf{x}_{\perp}\left|E\left(\ell / 2, \mathbf{x}_{\perp}\right)\right|^2-\left|E\left(-\ell / 2, \mathbf{x}_{\perp}\right)\right|^2.
\label{eqn:linear_x_int}
\end{split}
\end{align}
We now linearize the interaction \eqref{eqn:linear_x_int} around a coherent background laser. In particular, we decompose the right-moving modes a coherent beam plus vacuum fluctuations. The interaction potential can be separated into three types of terms
\begin{equation}
V=V_{R R}+2 V_{R L}+V_{L L},
\end{equation}
where R (L) subscripts denote the right (left) direction of propagation of the modes in the interaction. Concretely, for instance, we have
\begin{equation}
\begin{aligned}
 V_{RL}&\supset-\frac{\chi_\mathrm{e}}{2} x \int d^2 \mathbf{x}_{\perp} E_R\left(\ell / 2, \mathbf{x}_{\perp}\right) E_L\left(\ell / 2, \mathbf{x}_{\perp}\right) \\
& =\chi_\mathrm{e} \frac{x}{2(2\pi)^6} \int d^2 \mathbf{x}_{\perp} \int d^2 \mathbf{k}_{\perp}^{\prime} d^2 \mathbf{k}_{\perp} \int_0^{\infty} d k d k^{\prime} \sqrt{\omega_{\mathbf{k}}}\sqrt{\omega_{\mathbf{k}'}}  \\
& \times\left[e^{-i\left(k \ell / 2+\mathbf{k}_{\perp} \cdot \mathbf{x}_{\perp}\right)} a_{k, \mathbf{k}_{\perp}}-e^{i\left(k \ell / 2+\mathbf{k}_{\perp} \cdot \mathbf{x}_{\perp}\right)} a_{k, \mathbf{k}_{\perp}}^{\dagger}\right] \\
& \times\left[e^{i\left(k^{\prime} \ell / 2-\mathbf{k}_{\perp}^{\prime} \cdot \mathbf{x}_{\perp}\right)} b_{k^{\prime}, \mathbf{k}_{\perp}^{\prime}}-e^{-i\left(k^{\prime} \ell / 2-\mathbf{k}_{\perp}^{\prime} \cdot \mathbf{x}_{\perp}\right)} b_{k^{\prime}, \mathbf{k}_{\perp}^{\prime}}^{\dagger}\right] \\
& =\frac{\chi_\mathrm{e} x}{2(2 \pi)^4} \int d^2 \mathbf{k}_{\perp} \int_0^{\infty} d k d k^{\prime} \sqrt{\omega_{k \mathbf{k}_{\perp}}} \sqrt{\omega_{k^{\prime},- \mathbf{k}_{\perp}}}  \\
& \times\left[e^{-i\left(k-k^{\prime}\right) \ell / 2} a_{k, \mathbf{k}_{\perp}} b_{k^{\prime},-\mathbf{k}_{\perp}}-e^{-i\left(k+k^{\prime}\right) \ell / 2} a_{k, \mathbf{k}_{\perp}} b_{k^{\prime}, \mathbf{k}_{\perp}}^{\dagger}\right. \\
& \left.-e^{i\left(k+k^{\prime}\right) \ell / 2} a_{k, \mathbf{k}_{\perp}}^{\dagger} b_{k^{\prime}, \mathbf{k}_{\perp}}+e^{i\left(k-k^{\prime}\right) \ell / 2} a_{k, \mathbf{k}_{\perp}}^{\dagger} b_{k^{\prime},-\mathbf{k}_{\perp}}^{\dagger}\right],
\end{aligned}
\end{equation}
where in the third equality we have used the approximation that the slab is infinitely large. The full left-right interaction is then
\begin{align}
            &V_{RL}=\frac{\chi_\mathrm{e}}{2(2 \pi)^4} \int d^2 \mathbf{k}_{\perp} \int_0^{\infty} d k d k^{\prime}
            \sqrt{\omega_{k \mathbf{k}_{\perp}}} \sqrt{\omega_{k^{\prime},- \mathbf{k}_{\perp}}}
            \\\notag
           & \left[e^{-i\left(k-k^{\prime}\right) \ell / 2} a_{k, \mathbf{k}_{\perp}} b_{k^{\prime},-\mathbf{k}_{\perp}}-e^{-i\left(k+k^{\prime}\right) \ell / 2} a_{k, \mathbf{k}_{\perp}} b_{k^{\prime}, \mathbf{k}_{\perp}}^{\dagger}\right. \\\notag
&-\left.e^{i\left(k+k^{\prime}\right) \ell / 2} a_{k, \mathbf{k}_{\perp}}^{\dagger} b_{k^{\prime}, \mathbf{k}_{\perp}} +e^{i\left(k-k^{\prime}\right) \ell / 2} a_{k, \mathbf{k}_{\perp}}^{\dagger} b_{k^{\prime},-\mathbf{k}_{\perp}}^{\dagger}\right] \\ \notag
&-(l\rightarrow(-l))
\end{align}
We now make the rotating wave approximation (note that now we're working in the Schrodinger picture) and drop the counter-rotating terms
\begin{align}
\begin{split}
& V_{R L}=- i \frac{\chi_\mathrm{e}}{(2 \pi)^4} \\
&\times \int d^2 \mathbf{k}_{\perp} \int_0^{\infty} d k d k^{\prime} \sqrt{\omega_{k \mathbf{k}_{\perp}}} \sqrt{\omega_{k^{\prime},- \mathbf{k}_{\perp}}}  \\
& \times \sin \left[\left(k+k^{\prime}\right) \ell / 2\right]\left(a_{k, \mathbf{k}_{\perp}} b_{k^{\prime}, \mathbf{k}_{\perp}}^{\dagger}-a_{k, \mathbf{k}_{\perp}}^{\dagger} b_{k^{\prime}, \mathbf{k}_{\perp}}\right).
\end{split}
\end{align}
Similar expressions can be derived for $V_{RR}$ and $V_{LL}$.

Now, suppose that the laser comes in from the left port. This  displaces the modes into a coherent laser mode plus vacuum fluctuations: 
\begin{equation}
a_{k, \mathbf{k}_{\perp}} \rightarrow (2\pi)^3\alpha_0 e^{i \omega_0 t} \delta\left(k-k_0\right) \delta^2\left(\mathbf{k}_{\perp}\right)+a_{k, \mathbf{k}_{\perp}}.
\end{equation}
Note that $\alpha_0$ defined this way can be directly calculated by experimental parameters, given by $|\alpha_0|^2= \frac{I}{\omega}$, where $I$ is the intensity of light. We can define $
a_k\equiv\sqrt{\frac{1}{A}} a_{k, \mathbf{k}_{\perp}=0}, \quad b_k\equiv\sqrt{\frac{1}{A}} b_{k, \mathbf{k}_{\perp}=0}$, in terms of the laser cross-sectional area $A$. These operators 
 obey one-dimensional commutation relations: $\left[a_k, a_{k^{\prime}}^{\dagger}\right]=\delta\left(k-k^{\prime}\right)$.  By displacing the modes in this way, the amplitude of the right-moving modes is larger than the left-moving ones. To leading order in the large amplitude $\alpha_0$, the interaction Hamiltonian can be written as
 \begin{equation}
V_{R L}=x \int_0^{\infty} d k\left[g_k e^{i \omega_0 t} b_k^{\dagger}+g_k^* e^{-i \omega_0 t} b_k\right],
\end{equation}
where 
\begin{gather}
f_k =-i \frac{1}{2\pi}\alpha_0 \chi_\mathrm{e} \sqrt{A \omega_k \omega_0} \sin \left[\left(k_0-k\right) \ell / 2\right] \\
g_k =-i \frac{1}{2\pi}\alpha_0 \chi_\mathrm{e} \sqrt{ A\omega_k \omega_0} \sin \left[\left(k_0+k\right) \ell / 2\right].
\label{eqn:gk}
\end{gather}
This is our basic result for the interaction Hamiltonian in the presence of the laser drive, in the rotating wave approximation. The couplings are proportional to the square-root of the laser power ($\sim \alpha_0$), as usual in optomechanics.

By the same procedure, one can find $V_{LL}$ and $V_{RR}$. The RR term will contain a quadratic term in $\alpha_0$, a first-order term, and a zeroth-order term. The quadratic term will be given by $\frac{\chi_\mathrm{e}x}{2}|\alpha_0|^2$ and effects a change in the equilibrium of the slab, while the pure vacuum terms are subleading and will be omitted. Similarly, the $V_{LL}$ is subleading, and we will ignore it. The remaining non-trivial interaction term is then
\begin{equation}
 V_{R R}=x \int_0^{\infty} d k\left[f_k e^{i \omega_0 t} a_k^{\dagger}+f_k^* e^{-i \omega_0 t} a_k\right] \\
\end{equation}

Finally, the zeroth order term $V(0)$ in Eq.~\eqref{app-taylor} is an interaction term that is independent of the slab's center of mass coordinate:
\begin{equation}
    V(0)=-\frac{1}{2}\chi_e \int_{\text {slab }} d^3 \mathbf{r}|\mathbf{E}(\mathbf{x_0}, t)|^2.
\end{equation}
As with the interaction potential that was linear in $x$, we only want to keep the terms that are drive-enhanced. Let the the coherent laser light be $E_{\rm coh}$, and the right-moving and left-moving modes be as defined above \eqref{eq:lrdef}, then the dominant terms in the potential are given by
\begin{align}
\begin{split}
    &V(0)= \\
    &-\frac{1}{2} \chi_e \int_{\rm slab} d^3 r \, |E_{\rm coh}|^2+2 \mathbf{Re}(E_{\rm coh} E_R^\ast)+ 2\mathbf{Re}(E_{\rm coh} E_L^\ast).
\end{split}
\end{align}
The first term here is an irrelevant constant shift to the potential. The second term, meanwhile, can be evaluated within the rotating wave approximation as 
\begin{align}
   & \chi_\mathrm{e} \int d^2x_{\perp} dx \mathcal{E}_{k_0}\mathcal{E}_k(\alpha_0 e^{i\omega_0 t} e^{-ikx}-\alpha_0^\ast e^{-i\omega_0 t}e^{ikx}) \notag\\ &\int dk d^2 k_{\perp} (e^{-i (kx+k_\perp r_\perp)}a_{k,k_\perp}-e^{-i (kx_0+k_\perp r_\perp)}a_{k,k_\perp}) \notag
\\
    & \approx \chi_\mathrm{e}\int dk dx_0 \mathcal{E}_{k_0}\mathcal{E}_k\alpha_0( e^{i\omega_0 t} a_k^\dagger e^{i (k-k_0) x}+ e^{-i\omega_0 t} a_k e^{-i (k-k_0) x}) \notag\\
\end{align}
Integrating $x_0$ from $-\frac{l}{2}$ to $\frac{l}{2}$ , and one finds the potential
\begin{equation}
    V_0= \int dk  \frac{i}{k-k_0} (f_k e^{i\omega_0 t}a_k^{\dagger}-f_k^\ast e^{-i\omega_0 t}a_k).
\end{equation}
The potential for the left-moving modes has the same form. 

To summarize our results, the full interaction Hamiltonian of the slab and light is
\begin{align}
V(x) = V_{RR}(x) + 2 V_{RL}(x)+V(0),
\end{align}
with
\begin{align}
V_{R R}&=x \int_0^{\infty} d k\left[f_k e^{i \omega_0 t} a_k^{\dagger}+f_k^* e^{-i \omega_0 t} a_k\right], \\
 V_{R L}&=x \int_0^{\infty} d k\left[g_k e^{i \omega_0 t} b_k^{\dagger}+g_k^* e^{-i \omega_0 t} b_k\right], \\
    V(0) &= i\int dk \frac{1}{k-k_0} (f_k e^{i\omega_0 t}a_k^{\dagger}-f_k^\ast e^{-i\omega_0 t}a_k) 
\\    
    &+i \int dk\frac{1}{k+k_0} (g_k e^{i\omega_0 t}b_k^{\dagger}-g_k^\ast e^{-i\omega_0 t}b_k)
\end{align}

In the frame co-rotating with the laser beam, the equations of motion of the light are
\begin{equation}
\begin{gathered}
\dot{a}_k=-i \Delta_k a_k+i f_k x - \frac{f_k}{k-k_0} \\
\dot{b}_k=-i \Delta_k b_k+i g_k x - \frac{g_k}{k+k_0}.
\label{eq:eom}
\end{gathered}
\end{equation}
With these in hand, we also obtain the equation of motion for the mechanics
\begin{align}
\dot{x} & = p / m \\
\begin{split}
\dot{p} &=-m \omega_m^2 x -\int_0^{\infty} d k (f_k a_k^{\dagger}+f_k^* a_k ) \\
&-\int_0^{\infty} d k (g_k b_k^{\dagger}+g_k^* b_k). \label{eq:meom}
\end{split}
\end{align}
The solution of equation \eqref{eq:eom} can be written as 
\begin{align}
\begin{split}
    a_k(t)&=a^{\text {in }}_k(t) \\
    &+i \int_{t_0}^t d t^{\prime}  f_ke^{-i \Delta_k\left(t-t^{\prime}\right)} \left ( x(t^{\prime})+ i \frac{1}{k-k_0} \right), \label{eq:solution}
\end{split}
\end{align}
where  $a^{\text {in }}_k(t)  =e^{i \Delta_k (t-t_0)} a_k\left(t_0\right) $ is written in terms of an initial boundary condition at time $t_0$. Similarly, we can define $a^{\text {out }}_k(t) = e^{-i\Delta_k(t-t_1)}a_k(t_1)$ and write the solution in terms of late-time boundary conditions at $t_1$
\begin{align}
\begin{split}
    a_k(t)&=a^{\text {out }}_k(t)\\
    &-i  \int_{t}^{t_f} d t^{\prime} f_k e^{-i \Delta_k\left(t-t^{\prime}\right)} \left ( x(t^{\prime})+ i \frac{1}{k-k_0} \right),
\end{split}
\end{align}
The difference of these two equations gives 
\begin{align}
\begin{split}
    &a^{\text {out }}_k(t)-a^{\text {in }}_k(t)\\
    &= i  \int_{t_0}^{t_f} d t^{\prime}  f_ke^{-i \Delta_k\left(t-t^{\prime}\right)} \left ( x(t^{\prime})+ i \frac{1}{k-k_0} \right). 
\end{split}
\end{align}
This is the basic in-out relation for the optical field monitoring the slab.

For convenience, we now set $f$ and $g$ as real numbers, meaning we choose the phase of $\alpha_0$ so that it is imaginary. This choice of phase implies that the effect of changes in position are imprinted on the phase quadrature of the output light.  Converting to the phase and amplitude quadratures, 
\begin{align}
    X_k(t)&\equiv \frac{a_k+a_k^\dagger}{\sqrt{2}} \\
    &= X^{\rm in}_k+ \frac{ 1}{\sqrt{2}}\int_{t_0}^t d  t^{\prime}  \frac{f_k}{k-k_0}(e^{-i\Delta_k(t-t')}+e^{i\Delta_k(t-t')}),  \notag\\
    Y_k(t)&\equiv \frac{a_k-a_k^\dagger}{\sqrt{2}i}\\
    &=Y ^{\rm in}_k+\frac{ 1}{\sqrt{2}}\int_{t_0}^t d  t^{\prime}  f_kx(t')(e^{-i\Delta_k(t-t')}+e^{i\Delta_k(t-t')}). \notag
\end{align}
Now we can make the Markov approximation: when $\left|k-k_0\right| \gtrsim 1 / \tau$, where $\tau$ is the interaction time, any phases that rapidly oscillate will average to zero.  Since the phonons, or excitations of the harmonic potential, oscillate with a certain amplitude and frequency $\omega_m$, the phase will only pick up relevant modes around $k_0+\omega_m$ and $k_0-\omega_m$. When the mechanical frequency is much smaller than the laser frequency, we can make an approximation
$f_k \equiv f=f_{k_0}$ and $g_k \equiv g=g_{k_0}$. Explicitly, by comparison with \eqref{eqn:gk}, we get
\begin{align}
    g = \frac{|\alpha_0| \chi_\mathrm{e}\omega_0}{2\pi} \sqrt{A} \sin k \ell.
    \label{eqn:coupling}
\end{align}

Then we can integrate over all modes and define $Y^{\text {in }}(t)  =\int_0^{\infty} d k Y_k^{\rm in} =\int_0^{\infty} d k \frac{a_k^{\text {in }}-a_k^{\text{in}\dagger}}{\sqrt{2}i} $ , to get the input-output equations
\begin{gather}
   Y_R^{\mathrm{out}}-Y_R^{\mathrm{in}}=\sqrt{2}f x, \\
   Y_L^{\mathrm{out}}-Y_L^{\mathrm{in}}=\sqrt{2}gx.
\end{gather}
As anticipated, the the center-of-mass position has been imprinted on the phase of the output light. The amplitude quadratures of the light will instead be shifted coherently, and obey different in-out relations
\begin{align}
    X_R^{\mathrm{out}}-X_R^{\mathrm{in}}& = \sqrt{2 \Gamma_R}\\
   X_L^{\mathrm{out}}-X_L^{\mathrm{in}} & =\sqrt{2\Gamma_L},
\end{align}
for both left and right-moving modes. Here we defined $\sqrt{\Gamma_R}=\frac{f_k}{k-k_0} |_{k=k_0}=-i\frac{k_0\ell}{4}\alpha_0 \chi_\mathrm{e} \sqrt{A} $ and $\sqrt{\Gamma_L}=\frac{g_{k_0}}{2k_0}=-i \frac{1}{4}\alpha_0 \chi_\mathrm{e} \sqrt{A } \sin \left(k_0\ell \right)$, and used
\begin{align}
\begin{split}
\int_0^{\infty} dk e^{-i\left(k-k_0\right)\left(t-t^{\prime}\right)} &\approx \int_{-\infty}^{\infty} d k e^{-i\left(k-k_0\right)\left(t-t^{\prime}\right)}\\
&=2 \pi \delta\left(t-t^{\prime}\right).
\end{split}
\end{align}

We now turn to solving for the dynamics of the slab. Plugging Eq.~\eqref{eq:solution} into Eq.~\eqref{eq:eom}, we find the a single second-order differential equation for the center-of-mass position 
\begin{equation}
\begin{split}
     m\ddot{x} + m \omega^2 x = -\sqrt{2}f X_R^{\tin} - \sqrt{2} g X_L^{\tin} +F^{\tin}+ \\\int_0^{\infty} dk \left (\frac{|f_k|^2}{k-k_0}+\frac{|g_k|^2}{k+k_0} \right ) 4\pi \delta(k-k_0).  \label{eq:eomx} 
\end{split}
\end{equation}
Here we have used $ \int_{t_0}^tdt' (e^{i\Delta_k (t-t')}+e^{-i\Delta_k (t-t')}) =4\pi \delta(k-k_0) $, and defined the mechanical
susceptibility $\chi_m^{-1}[\Omega]= m( -\Omega^2 + \omega_m^2 ) $. After Fourier transformation, we get
\begin{align}
    Y^{\tout}_L = Y^{\tin}_L + \sqrt{2} g \chi_m \big (& -\sqrt{2} f X_R^{\tin} - \sqrt{2} g X_L^{\tin} + \notag\\ F^{\tin}&+F_0 2\pi \delta(\Omega)  \big ) \\
     Y^{\tout}_R = Y^{\tin}_R + \sqrt{2} f \chi_m \big (& -\sqrt{2} f X_R^{\tin} - \sqrt{2} g X_L^{\tin} +\notag\\ F^{\tin} &+F_0 2\pi \delta(\Omega) \big ). 
     \label{eqn:slab-output}
\end{align}
These equations describe the behaviour of the scattered light $Y^{\tout}_R$ and the back-scattered light $Y^{\tout}_L$. The term $F_0$ is the last term in Eq. \eqref{eq:eomx}, which is a constant term, quadratic in the amplitude $|\alpha_0|$ -- it can be interpreted as a constant radiation pressure on the slab. 

Under the Markovian approximation, $f=0$, which implies that there is no information in the forward scattered light about the slab's position.\footnote{We can also consider the forward-scattered light. If we use this light to estimate the force, it has a noise PSD
\begin{gather}
  \begin{split}
        S_{FF,R}^{\rm out} = \frac{S_{YY,R}^{\rm in}}{2f^2|\chi_m|^2} + 2 g^2S_{XX,L}^{\rm in} +  2 f^2S_{XX,R}^{\rm in} \\+ \frac{2}{|\chi_m|^2}\mathbf{Re}\left (\chi_m S_{XY,R}^{\rm in} \right ).
  \end{split}
\end{gather}
For the 1D case, in the Markovian approximation, we have $f=0$, which implies that the SNR = 0, i.e., the forward scattered light contains no information about the force. The situation changes in the 3D case, which will have a similar form except that $f$ will be a non-zero number. Because of the scattering of light off of the forward axis, some of the position information will be encoded in the forward scattering. In this case, injecting squeeze light from the right port will be able to help reducing the quantum noise.} We thus focus on the back-scattered light, for which the PSD is 
\begin{equation}
  \begin{split}
        S_{YY,L}^{\tout} = S_{YY,L}^{\tin} +2g^2|\chi_m|^2 S_{FF}+ 4 f^2g^2|\chi_m|^2S^\tin_{XX,R} 
 \\+  4 g^4|\chi_m|^2S_{XX,L}^{\tin} + 2\mathbf{Re}\left (2g^2\chi_m S_{XY,L}^{\tin} \right ).
  \end{split}
\end{equation}
We can give some intuition of the terms: $S_{YY,L}$ is the shot noise, while $S_{XX,L}$ is the back-action noise, proportional to the square of mechanical susceptibility. There are also correlation terms that may be non-vanishing if one uses squeezed or non-Gaussian states of light. $F_0$ is the radiation pressure force on the slab, which would shift Y by shifting the slab position, but it has no fluctuations and will not contribute to the noise. 

The estimator for the force PSD is given by
\begin{equation}
S_{F F}(\nu)=\frac{S_{Y Y}^{\text {out }}(\nu)}{\left|\chi_{Y F}(\nu)\right|^2}
\end{equation}
Therefore the noise in the signal can be expressed as such for a back-scattered readout
\begin{equation}
 \begin{split}
        S_{FF,L}^{out} = \frac{S_{YY,L}^{in}}{2g^2|\chi_m|^2} + 2 g^2S_{XX,L} +  2 f^2S_{XX,R} + \\\frac{2}{|\chi_m|^2}\mathbf{Re}\left (\chi_m S_{XY,L}^{in} \right ). \label{eq:fs}
 \end{split}
\end{equation}
When we set $f=0$ in the Markov approximation, this may be written in the form of \eqref{eqn:PSDdef} with
\begin{align}
   \chi_{YX} = 1, \chi_{YY} = -2 g^2 \chi_m, \chi_{YF} = \sqrt{2} g \chi_m.
   \label{eqn:slabChis}
\end{align}

\subsection{Mapping between cavity optomechanics and dielectric slab}
\label{app:map}
In this appendix, we show how the dynamics of light interacting with a lossy (large $\kappa$) optomechanical cavity may be mapped on to the interaction with a dielectric slab.

Recall Eqs.~\eqref{eq:Y-out-linear-system} and \eqref{eqn:slab-output} for output phase of light for a cavity and slab, respectively, which we repeat here
\begin{align}
    Y_{\mathrm{cav}}^\tout &= (1 + \kappa \chi_c) Y^\tin - 2 \kappa g^2 \chi_c^2 \chi_m \, X^\tin - \sqrt{2 \kappa g^2} \chi_c \chi_m F^\tin \nonumber \\
    Y_{L,\mathrm{slab}}^\tout &= Y_L^\tin  - 2 g'^2 \chi_m X_R^\tin + \sqrt{2} g' \chi_m F^\tin,
\end{align}
where we have made the Markov approximation ($f=0$) in the expression for the phase quadrature of the left-moving light $Y_{L,\mathrm{slab}}^\tout$ and have called the coupling $g'$ to avoid confusion. In the limit of a bad cavity, such that $\kappa \gg \omega_m, \gamma, \nu$, we may approximate 
\begin{align}
    \chi_c(\nu) \approx - 2 / \kappa,
\end{align}
and so the output light from the cavity satisfies
\begin{align}
    Y_\mathrm{cav}^\tout \approx - Y^\tin - \frac{8 g^2}{\kappa} \chi_m X^\tin + \sqrt{8 g^2}{\kappa} \chi_m F^\tin.
\end{align}

\section{Optimal squeezing angle}
In this appendix, we give explicit functions for the optimal squeezing angle $\theta_*(\nu)$, given in Eq.~\eqref{eqn:optimal_angle}, for an optomechanical cavity. 

Recall that we had
\begin{align}
    \tan \theta_*(\nu) = \frac{2 \mathrm{Re} \big[ \chi_{YX} \, \chi_{YY}^\ast \big]}{|\chi_{YX}|^2 - |\chi_{YY}|^2}.
\end{align} 
We note since 
\begin{align}
    \chi_{YY} = - \chi_c / \chi_c^*,
\end{align}
we see that 
\begin{align}
\chi_{YX} \chi_{YY}^* &= (- 2 \kappa g^2 \chi_c^2 \chi_m) \times (-\chi_c^* / \chi_c) \nonumber \\
&= 2 \kappa g^2 |\chi_c|^2 \chi_m \nonumber \\
&= |\chi_{YX}| \times \frac{\chi_m}{|\chi_m|}.
\end{align}
Furthermore, since $|\chi_{YY}| = 1$, the optimal angle reduces to
\begin{align}
    \tan \theta_* (\nu) = \frac{2}{|\chi_{YX}| - |\chi_{YX}|^{-1}} \, \frac{\mathrm{Re}[\chi_m]}{|\chi_m|}. 
\end{align}
This reproduces the results of \cite{kimble2001conversion} when working in the free mass limit ($\nu \gg \omega_m$).

We now consider the limit of a bad-cavity -- such that $\chi_c(\nu) \approx -2/\kappa$ is flat in frequency -- and of a perfect mechanical element $\gamma = 0$. In this limit, the optimal angle is
\begin{align}
    \tan \theta_* (\nu) &\approx \frac{16 g^2 m \kappa (\omega_m^2 - \nu^2) }{ 64 g^4 - m^2 \kappa^2 (\omega_m^2 - \nu^2)^2  } \nonumber \\
    &= \frac{2}{\frac{8 g^2}{m \kappa (\omega_m^2 - \nu^2)}  - \frac{m \kappa (\nu^2 - \omega_m^2)}{8 g^2} },
\end{align}
and so, using
\begin{align}
    \tan 2x = \frac{2}{\cot x - \tan x},
\end{align}
we see that we need
\begin{align}
    \tan \frac{\theta_*(\nu)}{2} = \frac{m \kappa (\nu^2 - \omega_m^2)}{8 g^2}.
\end{align}

\newpage

\end{document}